\documentclass[letterpaper,11pt]{article}
\usepackage{jheppub}
\usepackage{slashed}
\usepackage{graphicx}
\usepackage{subfigure}
\usepackage{amsmath}
\usepackage{multirow}
\usepackage{comment}
\usepackage{hyperref}
\usepackage{epstopdf}
\usepackage{mathtools}

\newcommand{\beq}{\begin{equation}}
\newcommand{\eeq}{\end{equation}}
\newcommand{\bea}{\begin{eqnarray}}
\newcommand{\eea}{\end{eqnarray}}
\newcommand{\ba}{\begin{array}}
\newcommand{\ea}{\end{array}}

\def\m1{M_1}
\def\m2{M_2}
\def\m3{M_3}

\def\ni{{\chi^0_i}}

\def\ch10{\tilde \chi^0_1}

\def\gev{\,{\rm GeV}}

\newcommand{\lsim}{\mathrel{\mathop{\kern 0pt \rlap
  {\raise.2ex\hbox{$<$}}}
  \lower.9ex\hbox{\kern-.190em $\sim$}}}
\newcommand{\gsim}{\mathrel{\mathop{\kern 0pt \rlap
  {\raise.2ex\hbox{$>$}}}
  \lower.9ex\hbox{\kern-.190em $\sim$}}}

\definecolor{pink}{RGB}{255,105,180}

\newcommand{\ud}{\mathrm{d}}

\newcommand{\ifb}{{\,{\rm fb}^{-1}}}

\title{A potential scenario for the Majorana neutrino detection at future lepton colliders}

\author[a,b]{Yang Zhang}
\author[a,b]{, Bin Zhang}

\affiliation[a]{Department of Physics, Tsinghua University, Beijing, 100084, China}
\affiliation[b]{Center for High Energy Physics, Tsinghua University, Beijing, 100084, China}

\emailAdd{zhang-yang13@mails.tsinghua.edu.cn}

\abstract{The existence of Majorana neutrinos must lead to lepton-number violating processes, and the Majorana nature of neutrinos can only be experimentally verified only via  lepton-number violating processes. We propose a new approach to search for Majorana neutrinos at future electron-positron colliders by exploiting this feature. We investigate the $\Delta L = 2$ like-sign dilepton production and find that lepton colliders with different center-of-mass (c.m.) energies have comparative advantages in resonant production of a Majorana neutrino in either light neutrino mass range or heavy mass range. At the future Circular Electron-Positron Collider (CEPC), with 250 GeV c.m. energy and 5 ab$^{-1}$ integrated luminosity, we find that there could be more significant sensitivity for resonant production of a Majorana neutrino in the mass range of 5-80 GeV than previous results at LEP2 or LHC. At the 1 TeV ILC with 1 ab$^{-1}$ integrated luminosity, it has better sensitivity while the neutrino mass is larger than 250 GeV.}

\keywords{Majorana Neutrino, Seesaw Mechanism, Lepton Collider, CEPC, ILC}

\begin{document}
\maketitle
\flushbottom

\section{Introduction}

A multitude of neutrino oscillations evidences \cite{Barger:2003qi,Patrignani:2016xqp} from experiments with solar, atmospheric, reactor and accelerator neutrinos have shown that neutrinos have small non-zero masses, while the origin of their masses is still a mystery. Standard model (SM) itself, which merely incorporates left-handed neutrinos $\nu_L$ in SU(2)$_L$ gauge group doublets, is unable to generate neutrino mass without giving up gauge symmetry and renormalizability, so additional particles must be added to extend SM. The simplest scheme is to introduce $n$ right-handed SM singlet neutrino fields $N_{R}$. They can couple to three flavour neutrinos through Yukawa coupling which will be Dirac mass terms $m_D(\overline{\nu_L}N_R+\mathrm{h.c.})$ after gauge symmetry breaking. More interestingly, these gauge singlet neutrino fields are also allowed to couple to their own charge conjugate fields to form Majorana mass terms $M(\overline{N^c_L}N_R+\mathrm{h.c.})$. This is known as the type-I seesaw mechanism \cite{Minkowski:1977sc,Mohapatra:1979ia,Yanagida:1979as,GellMann:1980vs,Schechter:1980gr}, which has strong theoretical motivation and can be incorporated into plenty of scenarios, such as Left-Right symmetric gauge theories \cite{Pati:1974yy}; $SO(10)$ Supersymmetric (SUSY) grand unification \cite{Harvey:1981hk} and other grand unified theories \cite{Dorsner:2005fq}; models with exotic Higgs representations \cite{Zee:1985rj,Ma:1998dx}; R-parity violating interactions in Supersymmetry (SUSY) \cite{Barbier:2004ez} and theories with extra dimensions \cite{ArkaniHamed:1998vp}. Introduction of Majorana masses which are absent in SM and disobey the global U(1) symmetry, will result in lepton number no longer being conserved. Such unique phenomenon if observed, would in turn provide the most unambiguous evidence for existence of a Majorana neutrino. All those neutrino flavor oscillation phenomena imply mixture between the flavor and mass eigenstates of neutrinos. Therefore, via neutral current (NC) and charged current (CC), the direct collider search may produce lepton-number violating processes involving charged leptons helping to establish the nature of neutrino masses.

The prospects of detecting signals of heavy Majorana neutrinos on hadron colliders have been thoroughly explored \cite{Keung:1983uu,Dicus:1991fk,Almeida:2000pz,Panella:2001wq,Datta:1993nm,Bray:2007ru,delAguila:2007qnc,Han:2006ip,Atre:2009rg,Das:2012ze,Gluza:2015goa,Gluza:2016qqv,Guo:2017ybk}. Both ATLAS and CMS experiments at the LHC have set limits on the mixing parameters for heavy Majorana neutrino masses between 100 and 500 GeV \cite{Chatrchyan:2012fla,Khachatryan:2015gha,Aad:2015xaa}. The ``smoking gun'' signal at hadron colliders is the Majorana neutino production via Drell-Yan (DY) process: $pp\rightarrow W^{\ast}\rightarrow N\ell^\pm\rightarrow\ell^\pm \ell^\pm jj$. For the Majorana neutrinos which are lighter than $W$ bosons, the signal production rate is very large because they can come from on-shell $W$ bosons decay, while the jets in signals always have small transverse momentum in this case. There are complex backgrounds to be considered when detector effects are included. In order to discriminate against hadronic backgrounds at the LHC, more stringent acceptance cuts are applied which severely hurt the sensitivity of hadron coliders in the low mass region. Without such hadronic backgrounds, lepton colliders have significant advantages on the Majorana neutrino detection in the low mass region. Considering different ongoing projects of the next generation lepton colliders \cite{Behnke:2013xla,CEPC-SPPCStudyGroup:2015csa}, it is also worth examining the potential for the Majorana neutrino detection of lepton colliders which provide a much cleaner environment.

\section{Models and Mixing Parameters }

We begin our discussion with describing two basic models, to show how Majorana neutrinos fit into the big family photo of SM particles.

First, let us see the simplest extension, the well known Type I seesaw with $n$ right-handed SM singlet neutrino $N_{aR}~(a = 1,2,...,n)$. After gauge symmetry breaking, there are Dirac mass terms coming from the normal Yukawa couplings. Together with Majorana mass terms, they consist of the full neutrino mass terms which can be written as \cite{Atre:2009rg}
\begin{align}
-\mathcal{L}^\nu_m & =\frac{1}{2}\left(\sum^3_{a=1}\sum^n_{b=1}(\overline{\nu_{aL}}m^\nu_{ab}N_{bR}+
\overline{N^c_{bL}}m^{\nu\ast}_{ba}\nu^c_{aR})\right.\notag\\
& \left.+\sum^n_{b,b^\prime=1}
\overline{N^c_{bL}}B_{bb^\prime}N_{b^{\prime}R}\right)+\mathrm{h.c.}
\end{align}
where $\nu_{aL} (a = 1,2,3)$ are three generations of flavor neutrinos defined dynamically with respect to charged leptons. After diagonalization, they transform into
\begin{equation}
-\mathcal{L}^\nu_m =\frac{1}{2}\left(\sum^3_{m=1}m_{\nu_m}\overline{\nu_{mL}}\nu^c_{mR}
+\sum^{3+n}_{m^\prime=4}m_{N_{m^\prime}}\overline{N^c_{m^\prime L}}N_{m^\prime R}\right)+\mathrm{h.c.}
\end{equation}
with the following mixing relations between flavor and mass eigenstates
\begin{align}
& \nu_{aL}= \sum^3_{m=1}U_{am}\nu_{mL}+\sum^{3+n}_{m^\prime=4}V_{am'}N^c_{m'L},\\
& UU^\dag+VV^\dag=I.
\end{align}
In terms of the mass eigenstates, the CC and NC interactions Lagrangian appears as
\begin{align}
\mathcal{L}&=-\frac{g}{\sqrt{2}}W^+_\mu\left(\sum^\tau_{\ell=e}\sum^3_{m=1}U^\ast_{\ell m}
\overline{\nu_m}\gamma^\mu P_L\ell\right) \notag\\
 & -\frac{g}{\sqrt{2}}W^+_\mu\left(\sum^\tau_{\ell=e}\sum^{3+n}_{m'=4}V^\ast_{\ell m'}
 \overline{N^c_{m'}}\gamma^\mu P_L\ell\right) \notag\\
 & -\frac{g}{2\cos_W}Z_\mu\left(\sum^\tau_{\ell=e}\sum^{3+n}_{m'=4}V^\ast_{\ell m'}
 \overline{N^c_{m'}}\gamma^\mu P_L \nu_\ell\right)+\mathrm{h.c.}
\end{align}
where $P_L$ is the left-handed chirality projection operator. As can be seen, $U_{\ell m}$ is the non-unitary version of the PMNS mixing matrix in our example, while $V_{\ell m}$ determines the weight of the heavy Majorana neutrinos in leptonic CC and NC interactions. The latter will be the dominant contribution to the lepton-number violation processes at colliders, since the masses of light neutrinos, namely $\nu_m (m = 1, 2, 3)$ are at most $\mathcal{O}$(eV) \cite{Atre:2009rg}. Moreover, the existence of $V_{\ell m}$ is also relevant to many other intriguing topics, such as nonunitarity of the light neutrino mixing matrix as well as lepton flavor violation (LFV) \cite{Antusch:2006vwa,Abada:2007ux,Malinsky:2009gw,Malinsky:2009df,Dev:2009aw,Forero:2011pc,LalAwasthi:2011aa,Dinh:2012bp,Humbert:2015epa}, lepton nonuniversality and electroweak precision tests \cite{delAguila:2008pw,Akhmedov:2013hec,Basso:2013jka,Antusch:2014woa,Antusch:2015mia,Antusch:2016brq}. Therefore we can derive constraints on the heavy neutrino mass and the mixing elements $V_{\ell m}$ from experimental observations.

Now let us take a look at the decay of heavy Majorana neutrino. For Majorana neutrino heavier than Higgs boson $M_N>M_H$, the decay modes are to a $W$ or a $Z$ or a Higgs boson plus a corresponding SM lepton. The corresponding partial widths are proportional to $|V_{\ell m}|^2$ \cite{Banerjee:2015gca}:
\begin{align}
\Gamma(N\rightarrow\ell^-W^+)&=\frac{g^2}{64\pi}|V_{\ell N}|^2\frac{M^3_N}{M_W^2}\left(1-\frac{M_W^2}{M_N^2}\right)^2\left(1+2\frac{M_W^2}{M_N^2}\right)\\
\Gamma(N\rightarrow\nu_\ell Z)&=\frac{g^2}{128\pi}|V_{\ell N}|^2\frac{M^3_N}{M_W^2}\left(1-\frac{M_Z^2}{M_N^2}\right)^2\left(1+2\frac{M_Z^2}{M_N^2}\right)\\
\Gamma(N\rightarrow\nu_\ell H)&=\frac{g^2}{128\pi}|V_{\ell N}|^2\frac{M^3_N}{M_W^2}\left(1-\frac{M_H^2}{M_N^2}\right)^2
\end{align}
For Majorana neutrino lighter than $W$ boson $M_N<M_W$, it decays via charged and neutral current interactions to SM leptons plus pseudoscalar or vector mesons, a detailed list of which can be found in \cite{Atre:2009rg}. Adding all the partial decay widths, the total width of a heavy Majorana neutrino goes like \cite{Han:2006ip}
\begin{equation}
\Gamma_N\approx\left\{
\begin{array}{l}
    \sum\limits_\ell |V_{\ell N}|^2\dfrac{G_F M_N^3}{8}\quad {\rm for}\quad M_N>M_Z,M_H \\
    \\
    \sum\limits_\ell |V_{\ell N}|^2\dfrac{G_F^2 M_N^5}{10^3}\quad {\rm for}\quad M_N\ll M_W.
\end{array}\right.
\end{equation}
It is still quite narrow even for $M_N\sim 1$ TeV with small mixing angles. For $M_N\gg M_W,M_Z,M_H$, the decay branching ratios are Br$(\ell^-W^+)\approx$ Br$(\nu_\ell Z)\approx$ Br$(\nu_\ell H)\approx 25\%$.

Another way to incorporate Majorana neutrinos is through SU(2)$_L$ lepton triplets which is known as Type III Seesaw mechanism \cite{Foot:1988aq,Franceschini:2008pz,Cai:2017mow}. Here we employ triplet leptons $\Sigma_L$ in the representation (1,3,0) of the SM gauge group \cite{Cai:2017mow}:
\begin{equation}
\Sigma_L=\Sigma^a_L \sigma^a=\left(
 \begin{matrix}
   \Sigma^0_L/\sqrt{2} & \Sigma^+_L \\
   \Sigma^-_L & -\Sigma^0_L/\sqrt{2}
  \end{matrix}
\right),\quad\Sigma^\pm_L\equiv\frac{\Sigma^1_L\mp i\Sigma^2_L}{\sqrt{2}},\quad
\Sigma^0_L=\Sigma^3_L,
\end{equation}
where $\Sigma^\pm_L$ have electric charges $Q=\pm1$, and the $\sigma^a~(a = 1,2,3)$ are the usual Pauli matrices. The relevant Lagrangian for generating neutrinos masses is given by
\begin{equation}
\mathcal{L}=\frac{1}{2}{\rm Tr}[\overline{\Sigma_L} i\slashed{D}\Sigma_L]-\frac{M_\Sigma}{2}\overline{\Sigma^0_L}\Sigma^{0c}_R
-M_\Sigma\overline{\Sigma^-_L}\Sigma^{+c}_R
-Y_\Sigma\overline{L}\Sigma^c_R i\sigma^2 H^\ast+{\rm h.c.}
\end{equation}
where the first three are the kinetic and mass terms of the triplet, and the last one is its Yukawa coupling to the SM left-handed lepton doublet $L$ and Higgs doublet $H$. After gauge symmetry breaking, the Yukawa coupling will bring about mass mixing among the triplet and SM leptons which is key to our process discussed below. In addition to the mixing between charged triplet leptons $\Sigma_L^\pm$ and SM charged leptons $e,\mu,\tau$, which is special to Type III Seesaw mechanism, $\Sigma^{0c}_R$ has the same Yukawa coupling and mass term as the right-handed neutrino of Type I Seesaw above.

After unitary transformation, we transfer to mass basis and use $N$ and $E^\pm$ to denote mass eigenstates of triplet leptons, then the gauge eigenstates become
\begin{align}
&\Sigma^\pm=U E^\pm-\sqrt{2} V\ell_m^\pm, \quad \Sigma^0= U N-V\nu_m \\
&\ell^\pm=U \ell_m^\pm+\sqrt{2} V E^\pm, \quad \nu= U N+V\nu_m \\
&{\rm with} \quad |U|\sim\mathcal{O}(1), \quad |V|\sim\frac{Y_\Sigma v_0}{\sqrt{2}M_\Sigma}\ll 1.
\end{align}
The relevant gauge interaction Lagrangian in mass basis is
\begin{align}
\mathcal{L}&\ni -\overline{E}(eUA_\mu \gamma^\mu+g\cos\theta_W U Z_\mu\gamma^\mu)E-gU\overline{E}W^-_\mu \gamma^\mu N \notag \\
&- \frac{g}{2\cos\theta_W}Z_\mu\left(V\overline{N}\gamma^\mu P_R\nu+\sqrt{2} V\overline{E}\gamma^\mu P_R \ell\right) \notag \\
&- g W^+_\mu\left(V\overline{\nu}\gamma^\mu P_L E+\frac{1}{\sqrt{2}} V\overline{N}\gamma^\mu P_R \ell\right)+{\rm h.c.}
\end{align}
At tree level, the heavy leptons $N$ and $E^\pm$ are degenerate in mass, and we have \cite{Cai:2017mow}
\begin{align}
m_\nu\approx\frac{Y^2_\Sigma v^2_0}{2M_\Sigma}, \quad M_N\approx M_E\approx M_\Sigma.
\end{align}
However radiative corrections will create a mass split and make the charged $E^\pm$ slightly heavier than $N$. So the available decay channels of heavy Majorana $N$ here are the same as that of Type I Seesaw \cite{Franceschini:2008pz}.

One thing we would like to point out is that merely a ``canonical'' Seesaw mechanism is not likely to be the final answer to explain the reality about neutrino portal. Besides, the parameter space of the minimal Seesaw mechanisms is often severely restricted and unreachable to current experiments. For example, from the minimal, high-scale Type I Seesaw constructions, we can obtain such a relationship for mixing elements of Majorana neutrino \cite{Cai:2017mow},
\begin{equation}\label{mixing}
|V_{\ell N}|^2 \sim \frac{m_\nu}{M_N}.
\end{equation}
For a heavy neutrino mass of $M_N\sim 100$ GeV, this implies $|V_{\ell N}|^2 \sim 10^{-14}-10^{-12}$, well below any foreseeable experiments can reach. This calls for generalization of the minimal Seesaw models, which would give rise to greater phenomenology. Hybrid Seesaw models are one type of variants in which two or more ``canonical'' mechanisms are combined. For example, in Type I+II Hybrid Seesaw model, the light neutrino mass matrix $M_\nu$ is given by \cite{Cai:2017mow}
\begin{equation}
M_\nu=M_L-M_D M_N^{-1}M_D^T.
\end{equation}
Here, $M_D, M_N$ are respectively the Dirac and Majorana mass matrixes from Type I model, whereas $M_L$ comes from the Type II mechanism. In this scenario, sub-eV neutrino masses can arise not only from parametrically small Type I and II masses but additionally from an incomplete cancellation of the two terms. Hence $|V_{\ell N}|^2$ could break the relation in Eq.~(\ref{mixing}), and need not to be that small.

Since our goal is to see the detection capability of near future experiments, in our analysis in Section.~\ref{pheno}, we take a model-independent phenomenological approach. The heavy Majorana neutrino mass $M_N$ and mixing elements $|V_{\ell N}|$ are treated as free parameters, beyond restriction of derived relations from any specific models. So far, neutrinoless double-$\beta$ decay ($0\nu\beta\beta$) experiments have set the most stringent bound on mixing with electrons \cite{Elliott:2004hr,Rodejohann:2011mu}:
\begin{equation}
\sum_N \frac{|V_{eN}|^2}{m_N}<5\times10^{-8}\ \gev^{-1}.
\end{equation}
This makes it difficult for colliders to produce events involving electrons. So instead, we turn to focus on muon relevant signals and seek to improve the detecting sensitivity on $V_{\mu N}$. The most stringent constraint on mixing with muon from LEP2 experiments requires $|V_{\mu N}|^2 \lesssim 10^{-4}-10^{-5}$ for $ 5\ \gev < m_N < 80\ \gev$ \cite{Adriani:1992pq,Abreu:1996pa,Akrawy:1990zq}. For neutrinos heavier than $Z$ bosons, there are also plenty of researches based on electron colliders \cite{delAguila:1987nn,Buchmuller:1991tu,Djouadi:1993pe,Azuelos:1993qu,Gluza:1995js,Gluza:1997ts,Acciarri:1999qj,Achard:2001qv,Heister:2004wr}. However, due to the limitation of the center-of-mass (c.m.) energy of previous LEP2 experiments, the most stringent constraint for heavy Majorana neutrino masses between 100 and 500 GeV is from the LHC experiments based on the production via the DY process \cite{Chatrchyan:2012fla,Khachatryan:2015gha,Aad:2015xaa}. As for even higher scales, another production mechanism will come into play, which is the vector boson fusion (VBF) channel $W\gamma\rightarrow N\ell^\pm$, benefiting from its collinear logarithmic $t-$channel enhancement \cite{Dev:2013wba,Alva:2014gxa}. It dominates the DY mechanism for $m_N\sim 1 $ TeV (770 GeV) at the 14 TeV LHC (100 TeV VLHC) \cite{Alva:2014gxa}. A comprehensive discussion regarding the upper bounds on the mixing angles from all kinds of experimental (and prospective) studies can be found in \cite{Alonso:2012ji,Deppisch:2015qwa,Antusch:2016ejd,Cai:2017mow,Cvetic:2018elt}.

\section{Collider Signatures}\label{pheno}

\begin{figure}[htbp]
  \centering
  \begin{tabular}{cc}
  \includegraphics[width=0.4\textwidth,clip]{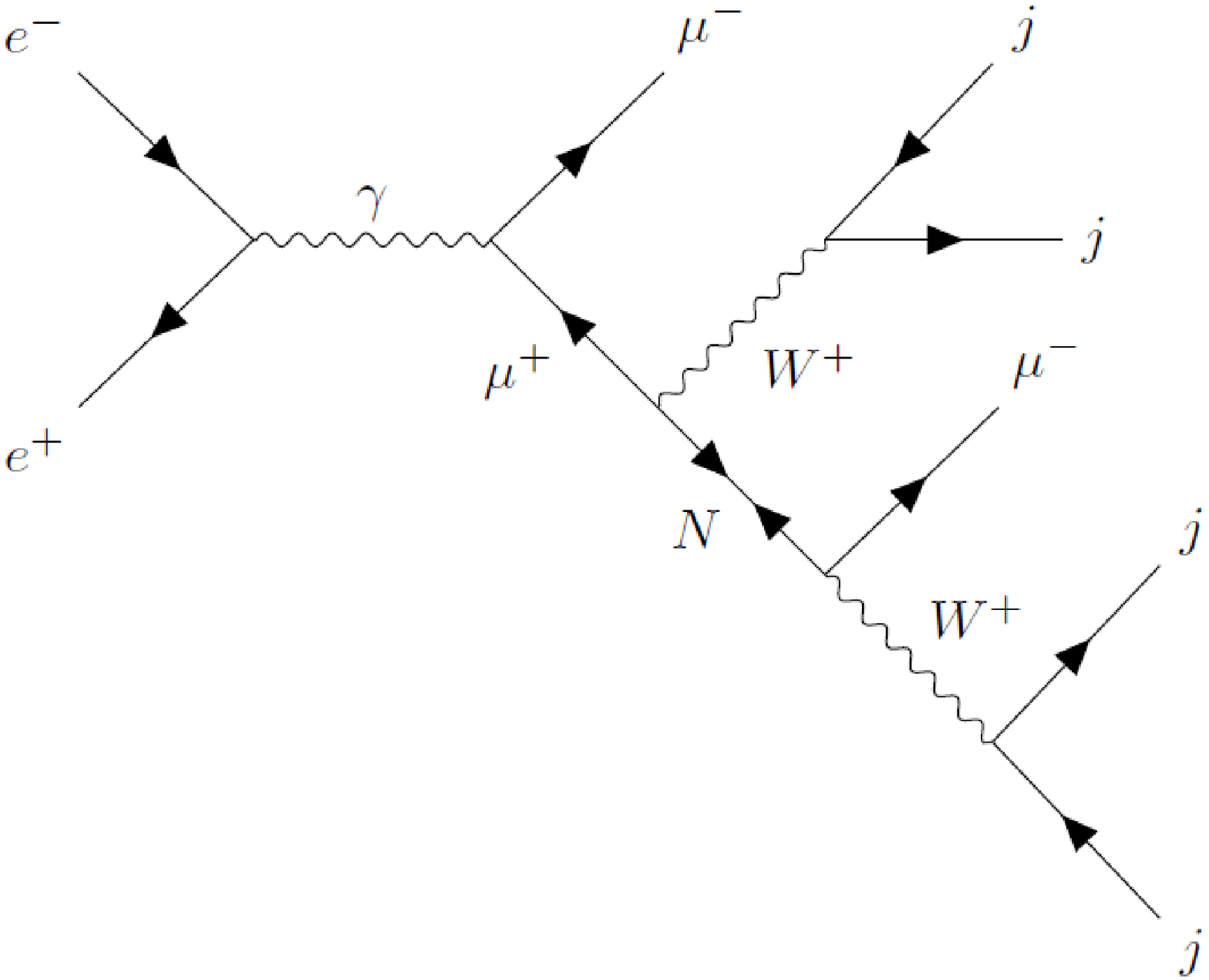}&
  \includegraphics[width=0.4\textwidth,clip]{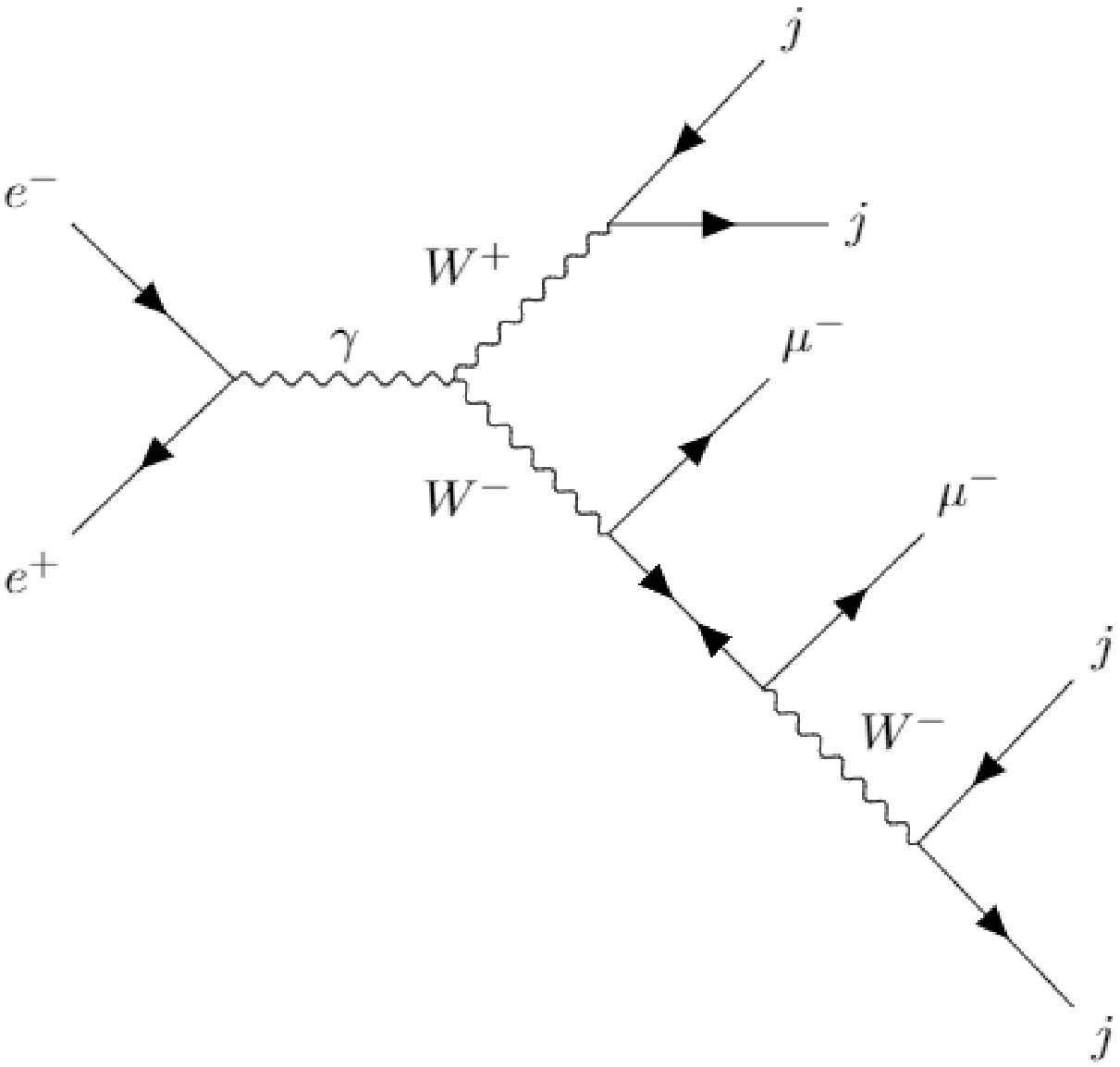}\\
  (a)&(b)\\
  \end{tabular}
  \includegraphics[width=0.4\textwidth,clip]{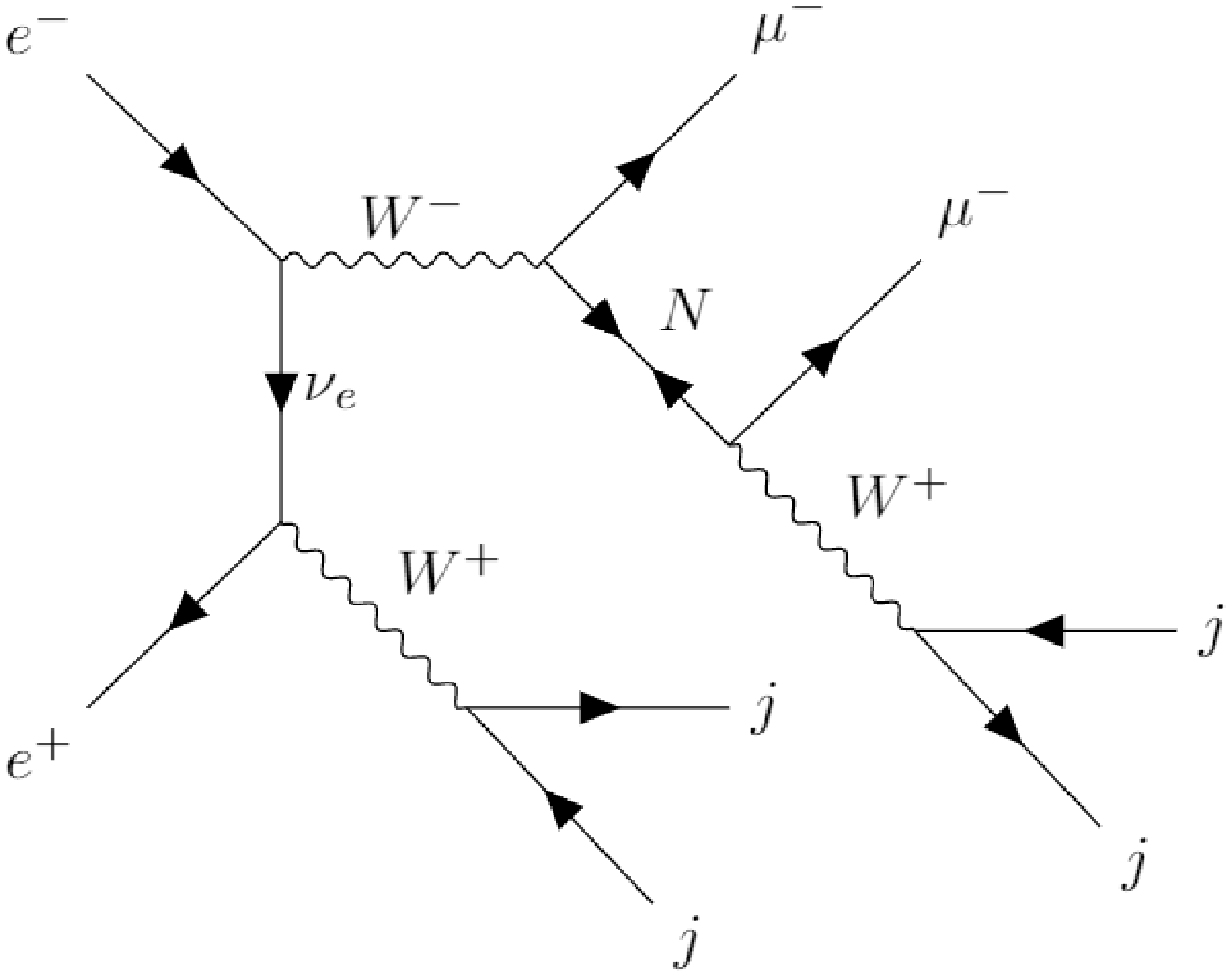}\\
  (c)
  \vspace{-0.3cm}
  \caption{Three sample Feynman diagrams for the process $e^+e^-\rightarrow N\mu^\pm W^\mp\rightarrow\mu^\pm\mu^\pm+4j$}
  \label{fig:Feyndiagram}
\end{figure}

\begin{figure}[tb]
\begin{center}
  \includegraphics[width=0.6\textwidth,clip=true]{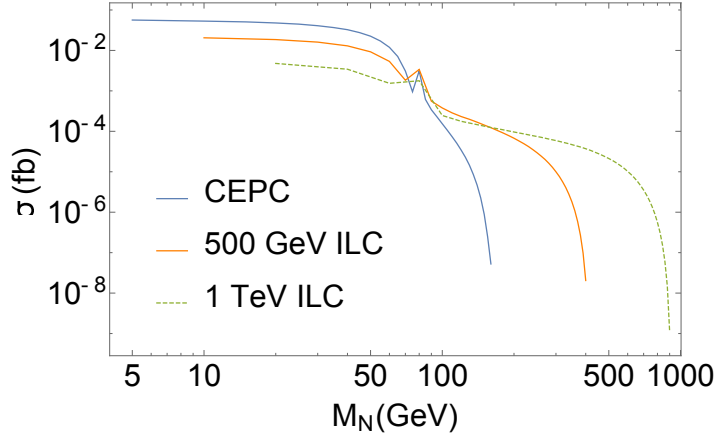}
  \caption{Total cross section for the process $e^+e^-\rightarrow N\mu^\pm W^\mp\rightarrow\mu^\pm\mu^\pm+4j$ at the CEPC and ILC (500 GeV and 1 TeV), with $S_{\mu\mu}=10^{-4}$.}
  \label{fig:rcs}
\end{center}
\end{figure}

The ``smoking gun'' signal for heavy Majorana neutrinos at hadron colliders is the like-sign dilepton final state with two jets and no missing transverse energy: $pp\rightarrow W^{\ast} \rightarrow N\ell^\pm\rightarrow\ell^\pm \ell^\pm jj$ \cite{Han:2006ip,Atre:2009rg}. While for electron-positron colliders, a similar approach has been proposed to produce lepton-number violation process, which is \cite{Banerjee:2015gca}
\begin{equation}
e^+e^-\rightarrow N\ell^\pm W^\mp\rightarrow\ell^\pm\ell^\pm+4j
\end{equation}
with same-sign dilepton plus four jets and no missing energy. For the reason mentioned above, we do not hold out much hope that electrons could play a significant role in this process. Hence we limit the dilepton signature in final states to be same-sign di-muon $\mu^\pm\mu^\pm$, and ignore mixing with electron $V_{eN}$ as well, giving the most conservative setting to derive bound for the value $V_{\mu N}$. Representative Feynman diagrams with di-muon in final states are displayed in Fig.~\ref{fig:Feyndiagram}.
Channels involving $eNW$ vertex have been omitted. In those diagrams, a Majorana neutrino can be produced on shell while kinematically accessible, and decays to a muon plus two jets soon afterwards. Due to the resonance enhancement effect, the total production rate is basically proportional to $|V_{\mu N}|^2$. For the Majorana neutrino mass range of our interest, the decay width is small enough so that the narrow-width approximation (NWA) is applicable. Thus the total cross section of our signal process can be approximately broken down as
\begin{align}\label{nwa}
 & \sigma(e^+ e^-\rightarrow \mu^\pm \mu^\pm +4j)\approx \notag \\
 & \sigma(e^+ e^-\rightarrow N\mu^\pm + 2j)\mathrm{Br}(N\rightarrow\mu^\pm+2j)\equiv S_{\mu\mu}\sigma_0
\end{align}
where $\sigma_0$ is basically a function of Majorana neutrino mass $m_N$, and independent of the mixing parameters when the heavy neutrino decay width is narrow. $S_{\mu\mu}$ is an ``effective mixing parameter'' of $N$ with a muon, and in the cases of minimal Type I and Type III Seesaw, can be further defined as \cite{Atre:2009rg}
\begin{equation}\label{emp}
S_{\mu\mu}=\frac{|V_{\mu N}|^4}{\sum^{\tau}_{\ell=e} |V_{\ell N}|^2}.
\end{equation}
This enables us to get a direct understanding of the effect of the mixing parameters. We calculate the cross sections of signal process at the CEPC and the ILC energies. In light of the possibility for a later upgrade of the ILC to (1 TeV), both 500 GeV and 1 TeV c.m. energies are considered. The total cross section values for different heavy neutrino masses are shown in Fig.~\ref{fig:rcs}, assuming a benchmark value for $S_{\mu\mu}=10^{-4}$.
This is the typical value of the bounds previously simulated at hadron colliders \cite{Atre:2009rg}. For $m_N<M_W$, the cross sections almost stay the same. While the mass approaches the c.m. energy limit, the cross sections drop sharply. The bumps at approximately 80 GeV correspond to the energy threshold of on-shell production of $W$ bosons in the decay of Majorana neutrinos. As we can see, the CEPC has an advantage in lower Majorana mass range (about 5-100 GeV) and the 1 TeV ILC in larger mass range (about 200-900 GeV). We will mainly present our results at 1 TeV for the ILC.

\subsection{Phenomenological Analysis at the CEPC}

Based on the discussion in the previous section, we study the Majorana neutrino production and lepton-number violating process at the future lepton colliders. The Circular Electron Positron Collider (CEPC) proposed by the Chinese high energy physics community is an ambitious project. Envisioned to operate at $\sqrt{s}\sim$ 250 GeV, CEPC may be able to shed light on Majorana neutrinos with lower mass ($m_N < 100$ GeV). Compared to linear colliders, it is easier for circular colliders to achieve a much higher integrated luminosity, which amounts to 5 ab$^{-1}$ in the case of CEPC. In this section, we take performance of the CEPC as benchmark for our numerical analysis \cite{CEPC-SPPCStudyGroup:2015csa}. To reach a high particle identification efficiency, we apply following basic trigger cuts on transverse momentum ($p_T$), pseudorapidity ($\eta$):
\begin{align}
 & p_T(\mu)>5~\gev,\quad\eta(\mu)<2.5,\label{cut1} \\
 & p_T(j)>5~\gev,\quad\eta(j)<2.5.\label{cut2}
\end{align}
To mimic the detector resolution effect, we also smear the four-momentum of the final state particles, following the performance specified in Ref.~\cite{CEPC-SPPCStudyGroup:2015csa}. For muons in the final state, the silicon tracking system provides high track momentum resolution, which performs approximately as
\begin{equation}\label{gs1}
\Delta\left(\frac{1}{p_T}\right)=2\times10^{-5}\oplus\frac{10^{-3}}{p_T\sin\theta}.
\end{equation}
The energy resolution of jets is determined by Hadron Calorimeter (HCAL) as
\begin{equation}\label{gs2}
\frac{\delta E}{E}=\frac{0.3}{\sqrt{E/\gev}}\oplus0.02.
\end{equation}

\begin{figure}[tbp]
  \centering
  \begin{tabular}{cc}
  \includegraphics[width=0.45\textwidth,clip]{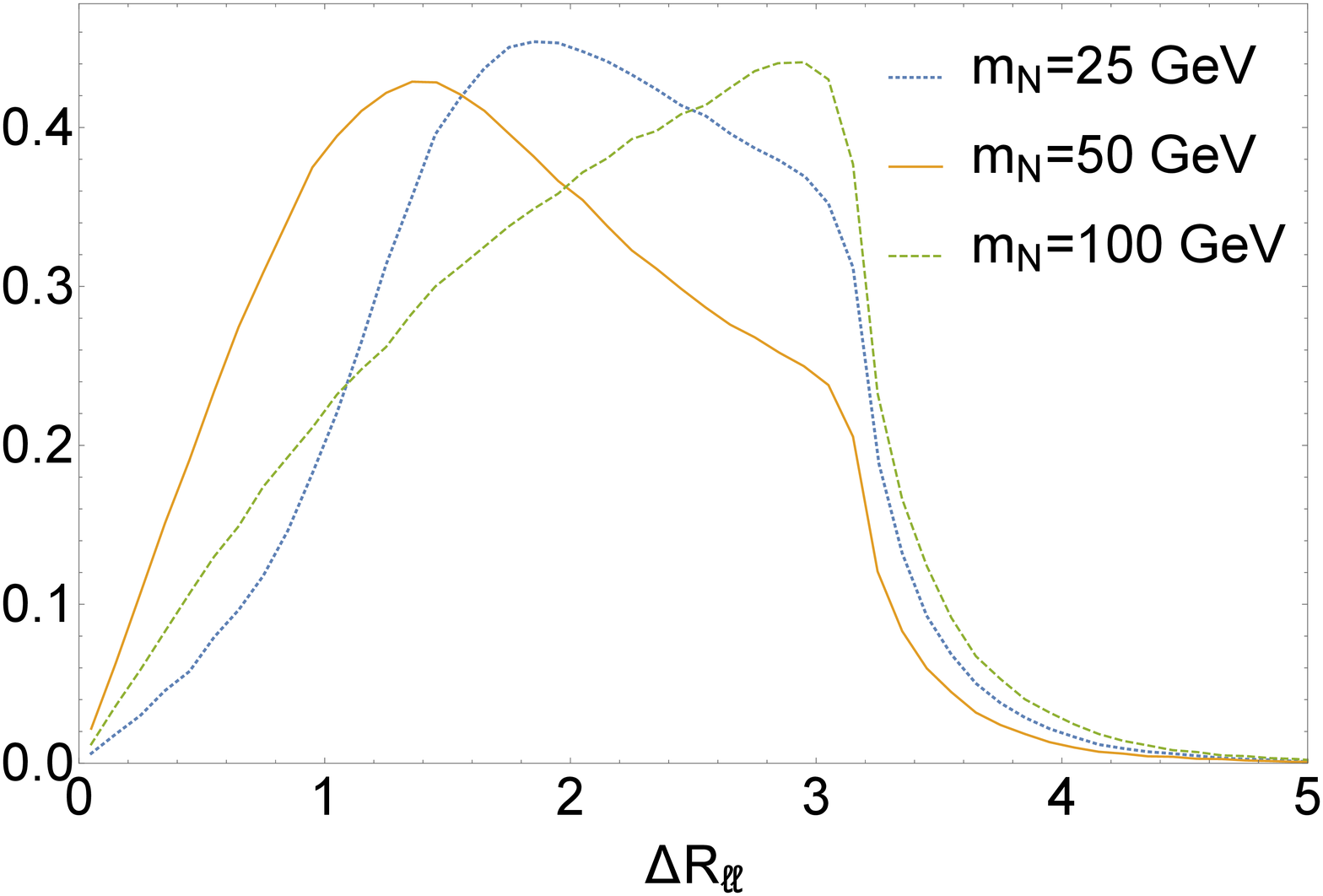}&
  \includegraphics[width=0.45\textwidth,clip]{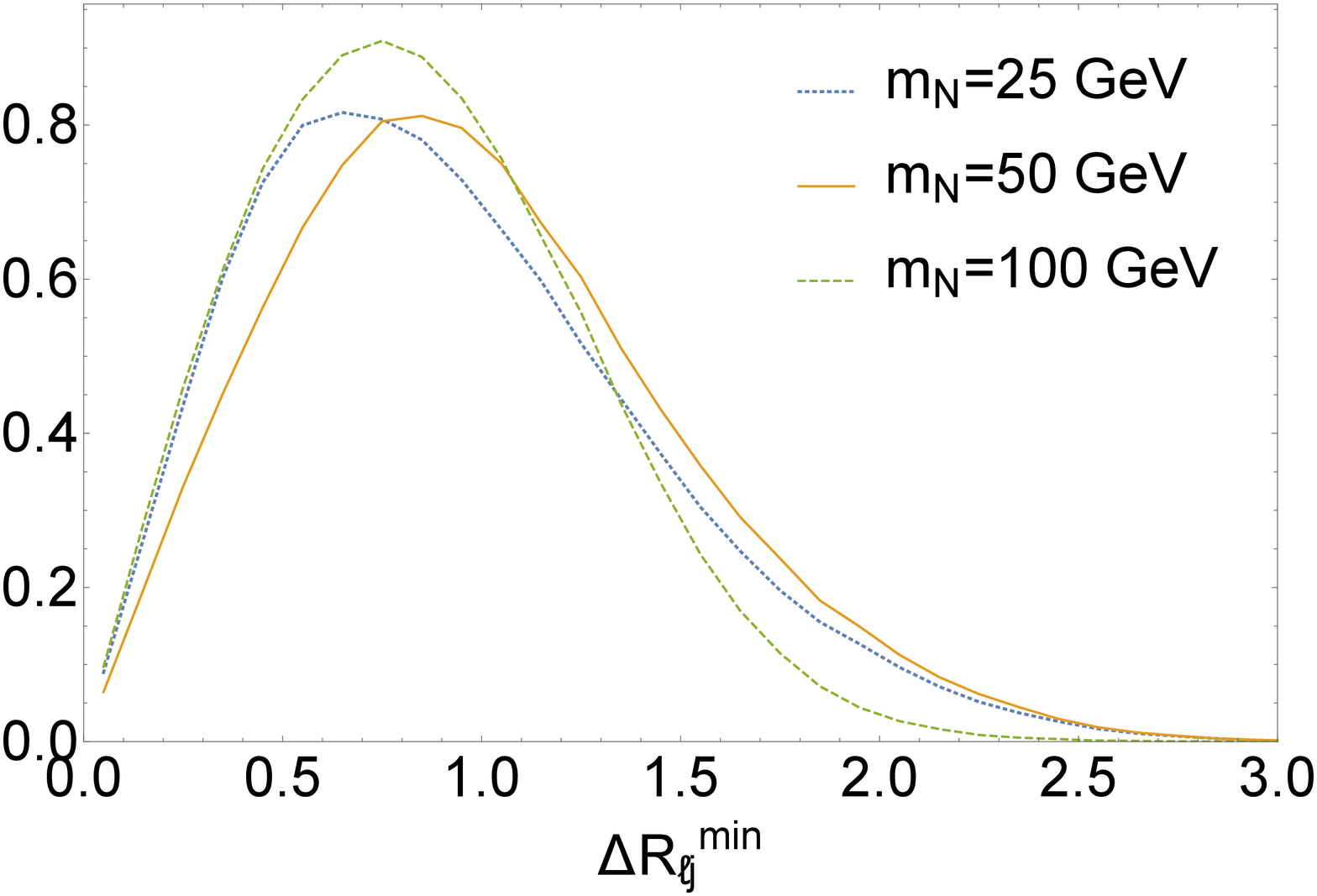}\\
  (a)&(b)\\
  \includegraphics[width=0.45\textwidth,clip]{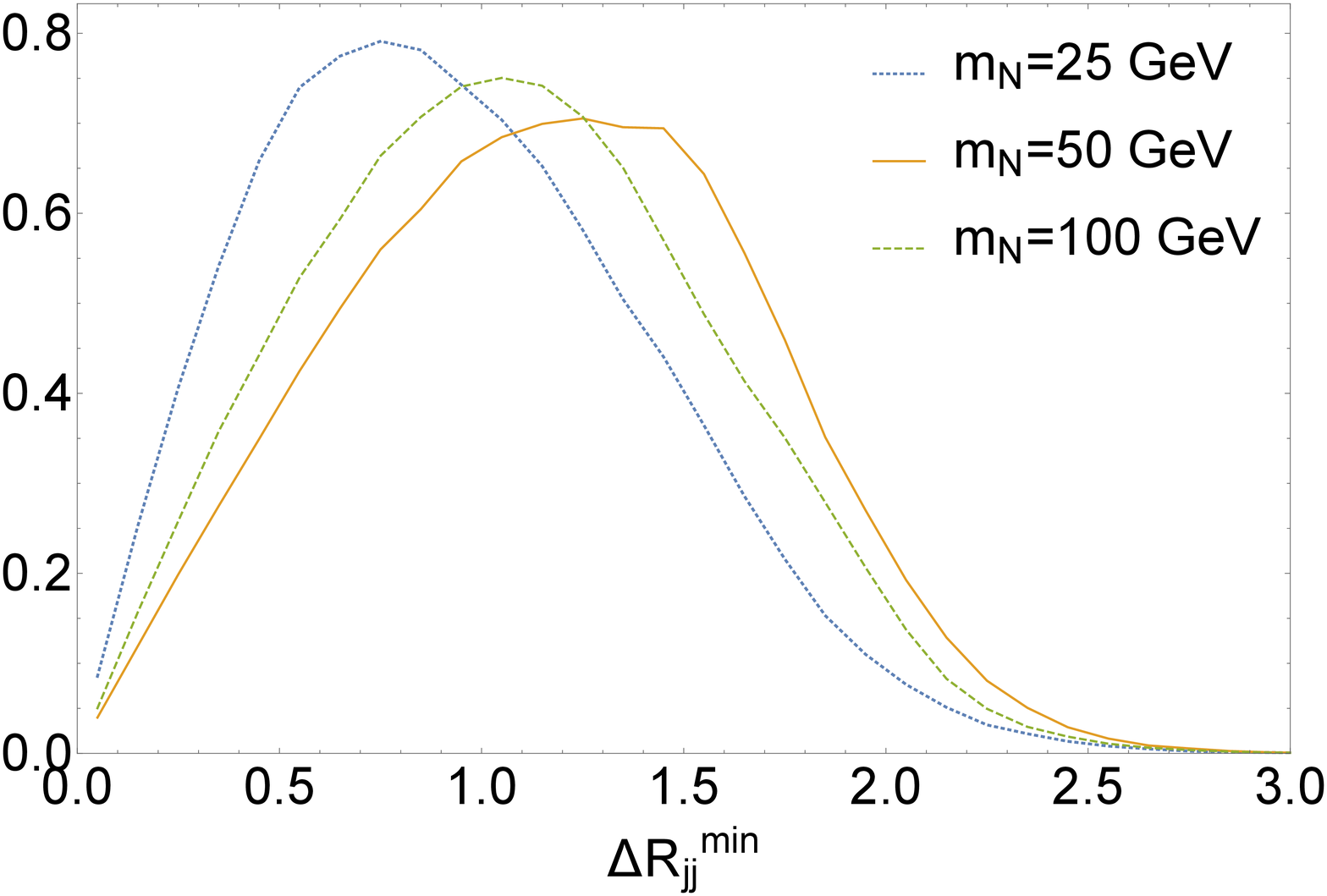}&
  \includegraphics[width=0.45\textwidth,clip]{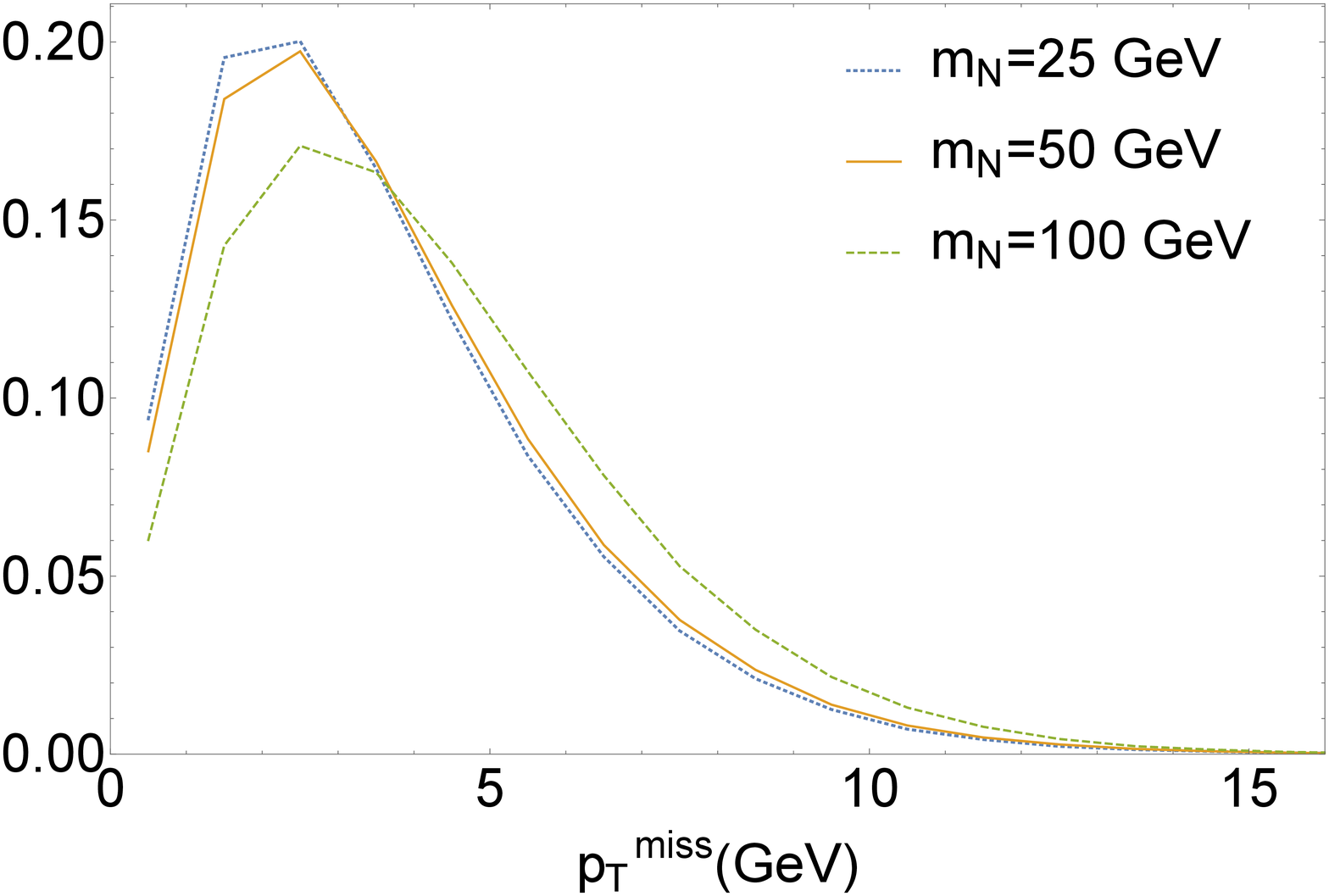}\\
  (c)&(d)\\
  \includegraphics[width=0.45\textwidth,clip]{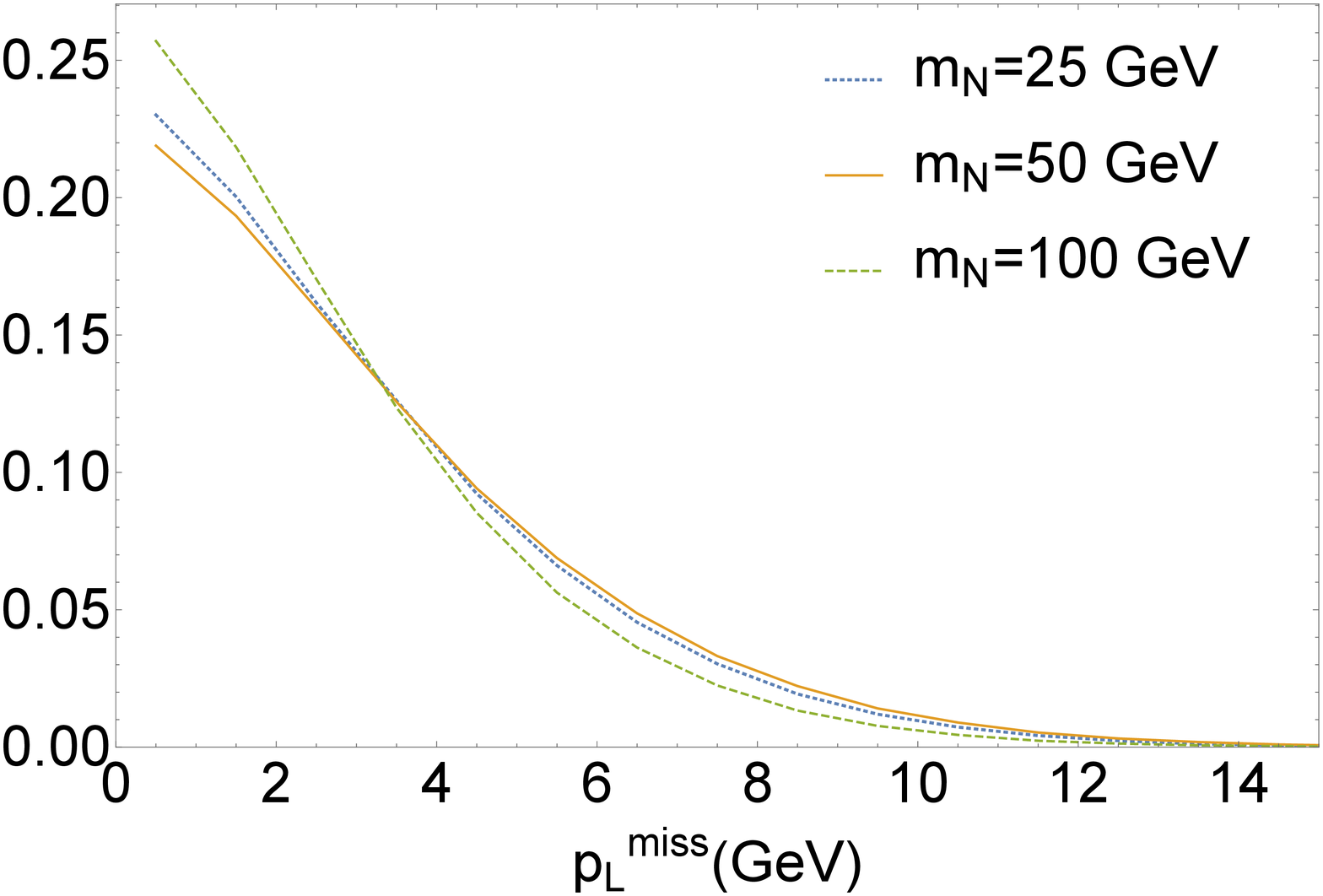}&
  \includegraphics[width=0.45\textwidth,clip]{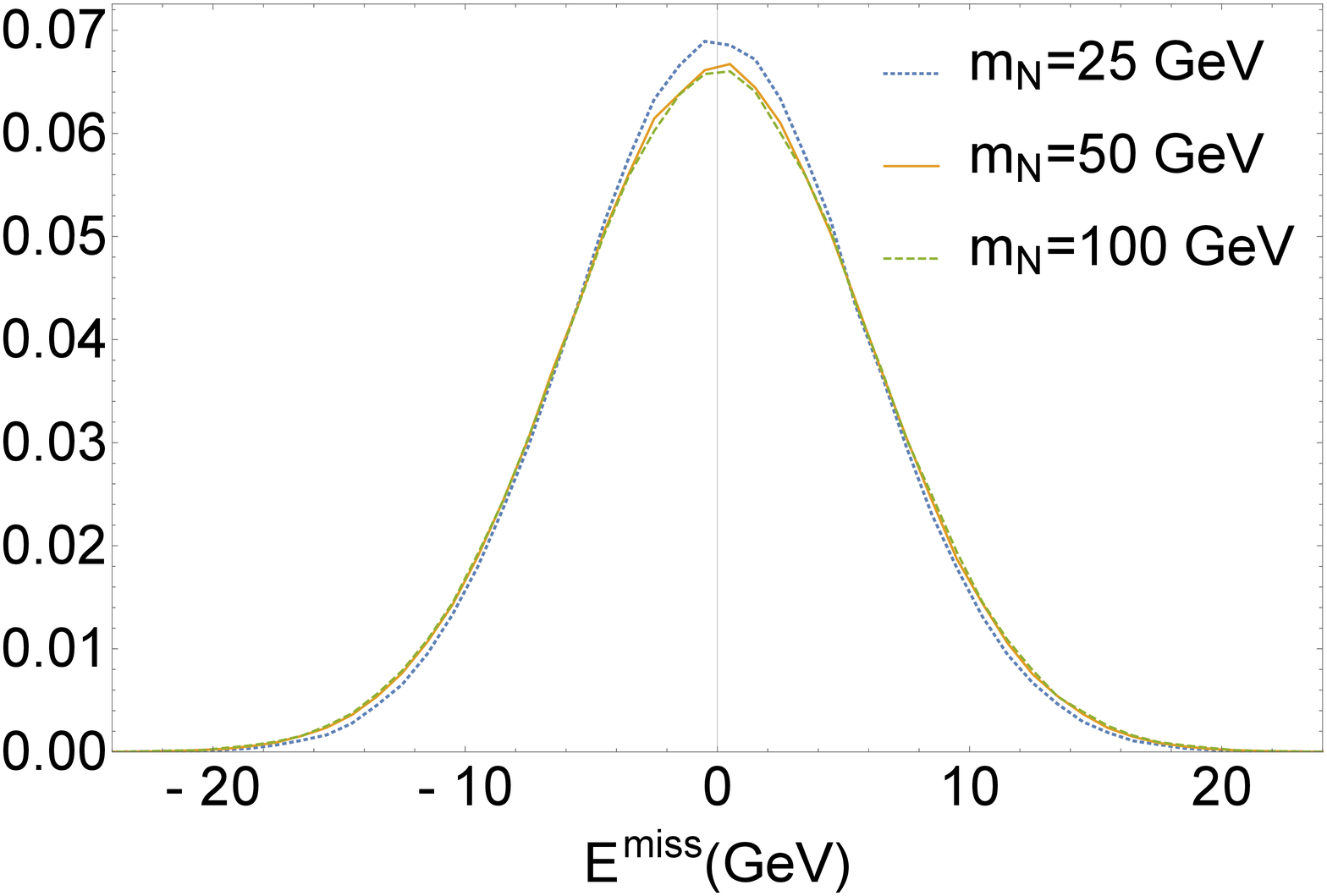}\\
  (e)&(f)
  \end{tabular}
  \caption{Normalized distributions $\sigma^{-1}\ud\sigma/\ud X$ for $m_N=25$, 50 and 100 GeV at the CEPC for (a) the radial isolation of muon pairs $\Delta R_{\ell\ell}$; the minimal radial isolation (b) $\Delta R_{\ell j}^{\rm min}$ and (c) $\Delta R_{jj}^{\rm min}$; (d) the missing transverse momentum $p\!\!\!\slash_T$; (e) the missing longitudinal momentum $p\!\!\!\slash_L$ and (f) the missing total energy $E\!\!\!\!\slash$. The missing total energy is c.m. energy 250 GeV minus the sum of energies of all visible particles in the final state.}
  \label{fig:cepcspc}
\end{figure}

To get a detailed understanding of our signal process, we further analysis the kinematical features of the final states. We choose three different masses $m_4=25$, 50 GeV (below $m_W$ threshold) and 100 GeV (above $m_W$ threshold) for illustration. The key point is there are two well-isolated same-sign muons, with no missing energy. Therefore we calculate the radial distance ($\Delta R=\sqrt{\Delta\eta^2+\Delta\phi^2}$) of different particle pairs in the final states: $\Delta R_{\ell\ell}$, $\Delta R_{\ell j}$ and $\Delta R_{jj}$ respectively. After normalization, the distributions of the minimal isolation are shown separately in Fig.~\ref{fig:cepcspc}(a), (b) and (c). Since all particles in the final state are visible in principle, by nature there is neither missing transverse momentum nor missing total energy in our signal process. However, misbalance in the energy-momentum measurement would still result from finite resolution of detectors, which is simulated by the smearing effect in Eqs.~(\ref{gs1})-(\ref{gs2}). This would be regarded as missing transverse momentum or missing energy. As an advantage over hadron colliders, the c.m. energy at an $e^+e^-$ collider is precisely measurable and adjustable. We exploit this feature by fully analyzing the missing transverse momentum, missing longitudinal momentum as well as missing total energy, and the distributions are plotted separately in Fig.~\ref{fig:cepcspc}(d), (e) and (f).
One more thing we would like to point out is the on-shell production of one Majorana neutrino $N$ and at least one $W^\pm$ boson in the signal process. Normally this could be useful for reconstructing them from invariant masses of final state particles, which are $m(\ell jj)$ in $N$ decay and $m(jj)$ in $W^\pm$ decay. However there are extra muon and jets from other sources which would impair this effect, and as we will see later, the missing momentum cuts already achieve great excellent discriminating power. Therefore the invariant mass cuts are not considered here.

\begin{figure}[tbp]
  \centering
  \includegraphics[width=0.5\textwidth,clip]{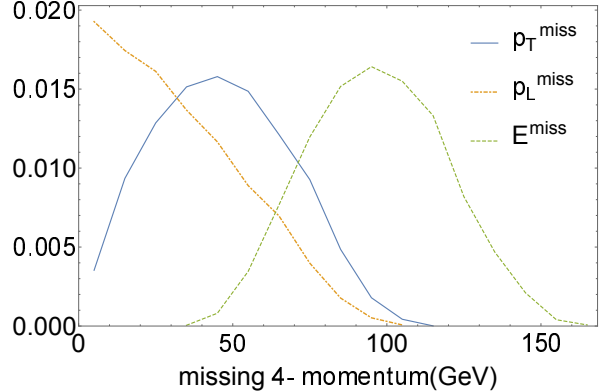}
  \caption{Normalized distributions $\sigma^{-1}\ud\sigma/\ud X$ for background channel $e^- e^+\rightarrow W^{+}Z+4j, W^{+}\rightarrow\mu^{+}+\nu_\mu, Z\rightarrow\mu^-\mu^+$ at the CEPC for the missing transverse momentum $p\!\!\!\slash_T$, missing longitudinal momentum $p\!\!\!\slash_L$ and missing total energy $E\!\!\!\!\slash$. The missing total energy is c.m. energy 250 GeV minus the sum of energies of all visible particles in the final state}
  \label{fig:mp250}
\end{figure}

Since SM Lagrangian conserves lepton number, there is no such process with $\Delta L=2$ in SM. It can only confuse the detectors' judgement by faking the like-sign dilepton signal when some particles are missing from detection. This comes from $4\ell+4j$ final states with two leptons lost in the beam pipe. By generating all processes with such final states in SM, we find most of them are at the level of $10^{-7}$ fb. What is left outstanding is the channel of diboson plus four jets: $e^- e^+\rightarrow WZ+4j$, with the two bosons decaying leptonically. We further analyze the kinematical information of this background channel and find the best discriminating power comes from missing four-momentum. Here, we display the distributions of $p\!\!\!\slash_T$, $p\!\!\!\slash_L$ and $E\!\!\!\!\slash$ of the background channel $e^- e^+\rightarrow W^{+}Z+4j, W^{+}\rightarrow\mu^{+}+\nu_\mu, Z\rightarrow\mu^-\mu^+$ in Fig.~\ref{fig:mp250}.
Combining the distributions of signal events, it is easy to see that $p\!\!\!\slash_T$ and $E\!\!\!\!\slash$ have the best discriminating powers, which could essentially eliminate this background.

\begin{figure}[tb]
\begin{center}
  \includegraphics[width=0.5\textwidth,clip=true]{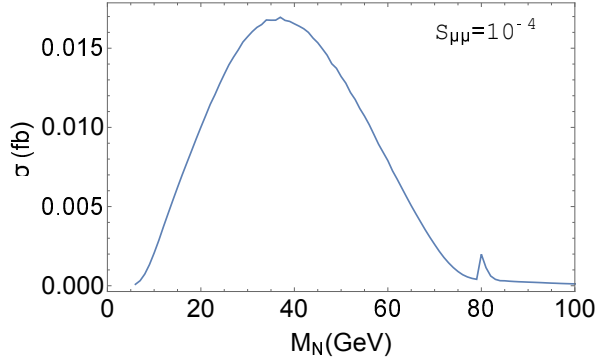}
  \caption{Total cross section for the process $e^+e^-\rightarrow N\mu^\pm W^\mp\rightarrow\mu^\pm\mu^\pm+4j$ after trigger cuts at the CEPC, with $S_{\mu\mu}=10^{-4}$.}
  \label{fig:cscut250}
\end{center}
\end{figure}

Based on the analysis hereinabove, we impose the following cuts to guarantee qualified events:
\begin{align}
 & \Delta R_{\ell\ell}>0.4,\quad \Delta R_{\ell j}^{\rm min}>0.4,\quad \Delta R_{jj}^{\rm min}>0.4,\label{cut3}\\
 & p\!\!\!\slash_T < 10~\gev,\quad E\!\!\!\!\slash < 20~\gev.\label{cut4}
\end{align}
After imposing the basic cuts in Eqs.~(\ref{cut1})-(\ref{cut2}) and these selection cuts in Eqs.~(\ref{cut3})-(\ref{cut4}), all these backgrounds are negligibly small. The signal cross sections for different heavy neutrino masses at the CEPC are shown in Fig.~\ref{fig:cscut250}, where we set $S_{\mu\mu}=10^{-4}$ as a benchmark value.
The reduction in rate is mainly due to basic acceptance cuts, especially in low mass range. The cross section reaches its maximum when neutrino mass is about 25 GeV, and then drops with increasing $m_N$. Although the cross section is only at the level of $10^{-2}-10^{-3}$ fb, the high integrated luminosity of the CEPC is able to make up for this shortage, bringing about enough events production.

\begin{figure}[tb]
\begin{center}
  \includegraphics[width=0.6\textwidth,clip]{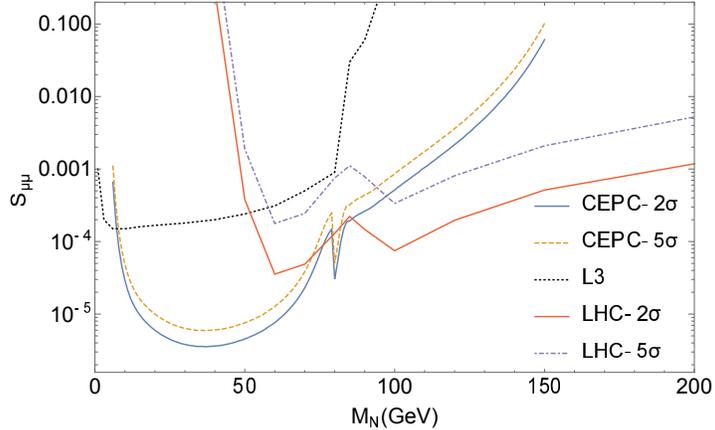}
  \caption{Sensitivity on $S_{\mu\mu}$ at the CEPC. For comparison, the 95\% bound from L3 search and simulated LHC results in \cite{Atre:2009rg} are presented.}
  \label{fig:excline250}
\end{center}
\end{figure}

To get signal significance, we consider Poisson statistics for the low event rate. For instance, a 95\% (2$\sigma$) bound on the signal for no background would need a signal event rate $N_S=\mathcal{L}\,\sigma_0 S_{\mu\mu}\geq 3$, where $\mathcal{L}$ is the integrated luminosity. Our final results about the sensitivity on the mixing parameter $S_{\mu\mu}$ are shown in Fig.~\ref{fig:excline250}, together with existing LEP limits from Z decay \cite{Adriani:1992pq} and the previous simulated LHC results in \cite{Atre:2009rg}.
The simulation results at the LHC are based on a c.m. energy of 14 TeV and an integrated luminosity of 100 $\ifb$. We see that there is significant improvement on detection sensitivity at the CEPC in the mass range of $m_N=5-80$ GeV. Compared to previous results, it improves measurement sensitivity on the mixing parameter $S_{\mu\mu}$ by one to two orders of magnitude in this mass range. In the mass range of $30-50$ GeV where has the best sensitivity, $S_{\mu\mu}$ can be probed to about $3\times10^{-6}$ at a $2\sigma$ level and $7\times10^{-6}$ at a $5\sigma$ level.

\subsection{Phenomenological Analysis at the ILC}

Since an $e^+e^-$ linear collider has the flexibility to operate at different energies and in different running conditions, experimental programs for the ILC are envisioned at series of
energies well adapted to individual physics goals. The exact run plan that will be carried out at the ILC will depend on the situation at the time of the ILC operation, taking into account new discoveries and measurements from the LHC running. For definiteness in our numerical simulation, we adopt a canonical program with c.m. energy of 1 TeV and integrated luminosity 1000 $\ifb$ \cite{Behnke:2013xla}. The proposed future lepton colliders in general have similar detector performance, therefore we apply the same basic acceptance cuts as that of CEPC on transverse momentum ($p_T$), pseudorapidity ($\eta$), as well as the same smearing effect, which are
\begin{align}
 & p_T(\mu)>5~\gev,\quad\eta(\mu)<2.5,\label{cut5} \\
 & p_T(j)>5~\gev,\quad\eta(j)<2.5.\label{cut6}\\
 & \Delta\left(\frac{1}{p_T}\right)=
 2\times10^{-5}\oplus\frac{10^{-3}}{p_T\sin\theta},\label{gs3}\\
 & \frac{\delta E}{E}=\frac{0.3}{\sqrt{E/\gev}}\oplus0.02.\label{gs4}
\end{align}

Following the similar approaches, we analysis the kinematical features of the final state particles and present their distributions. Since the ILC is capable of producing much more massive Majorana neutrinos, this time we choose $m_N=100$, 200 GeV and 500 GeV as examples for illustration. After normalization, the distributions of the minimal isolation of different particle pairs in the final state: $\Delta R_{\ell\ell}$, $\Delta R_{\ell j}^{\rm min}$ and $\Delta R_{jj}^{\rm min}$, are shown separately in Fig.~\ref{fig:ilcspc}(a), (b) and (c). And Fig.~\ref{fig:ilcspc}(d), (e) and (f) are respectively the distributions of the missing transverse momentum $p\!\!\!\slash_T$, missing longitudinal momentum $p\!\!\!\slash_L$ and missing total energy $E\!\!\!\!\slash$ incurred by finite resolution of the detectors.
\begin{figure}[tb]
  \centering
  \begin{tabular}{cc}
  \includegraphics[width=0.45\textwidth,clip]{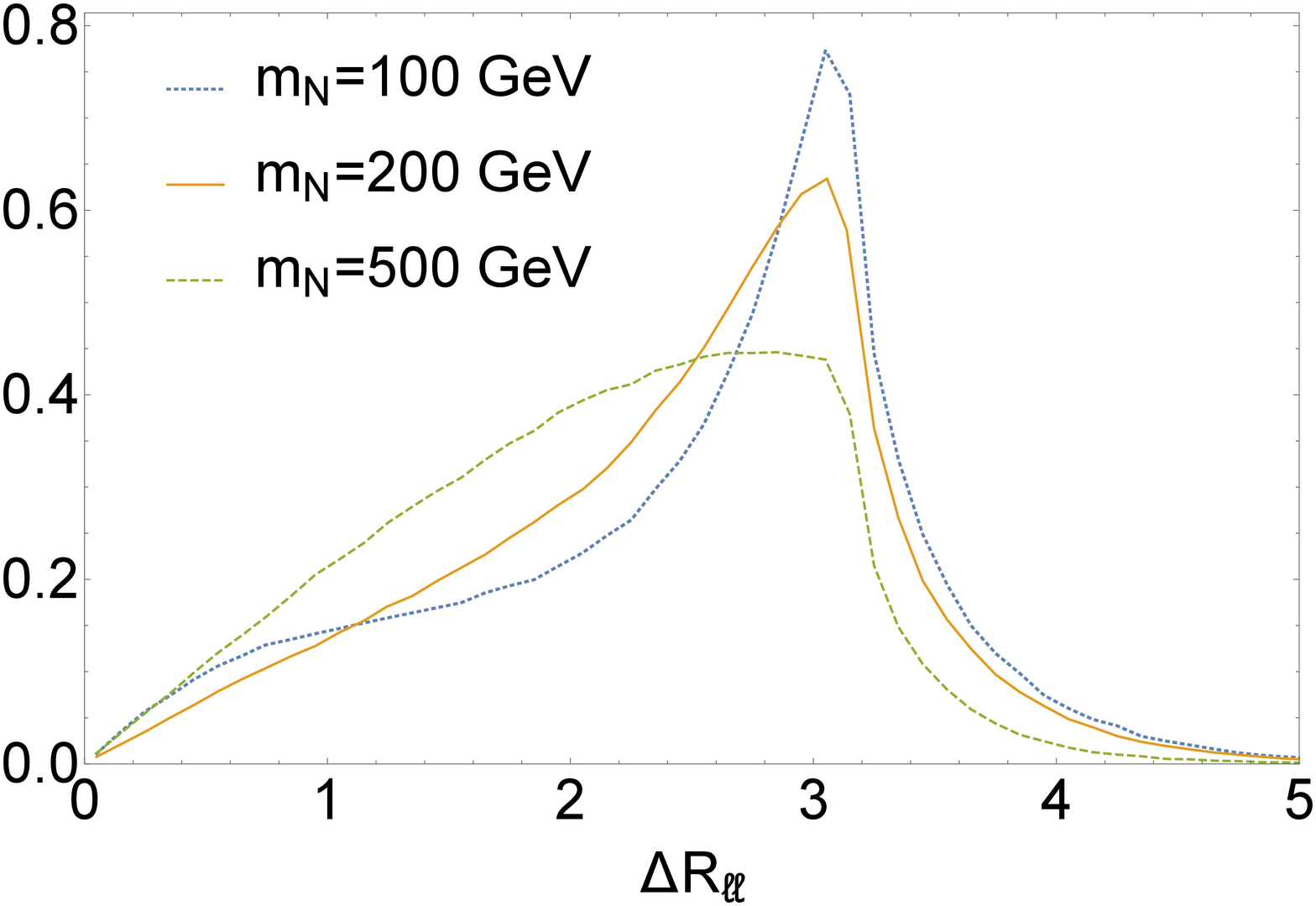}&
  \includegraphics[width=0.45\textwidth,clip]{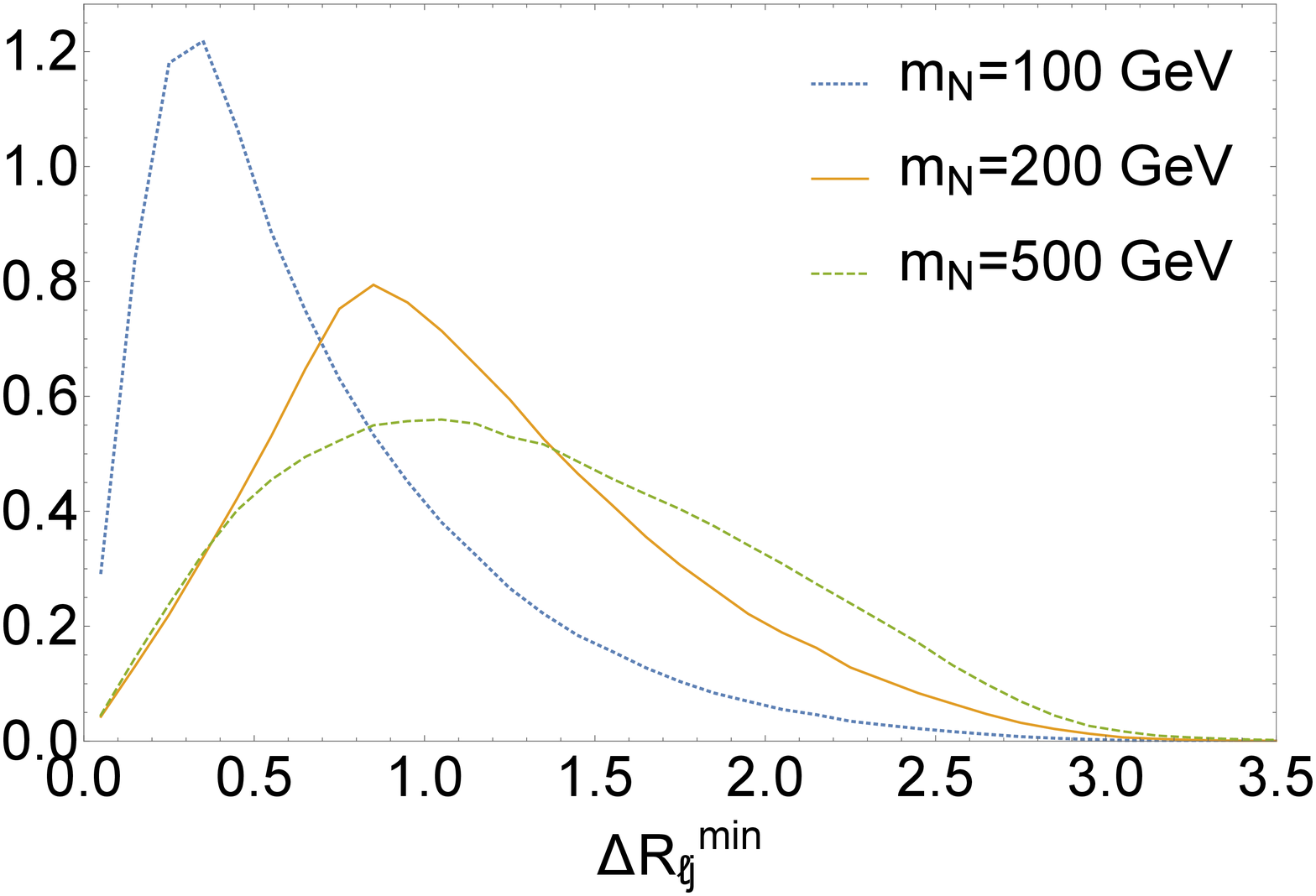}\\
  (a)&(b)\\
  \includegraphics[width=0.45\textwidth,clip]{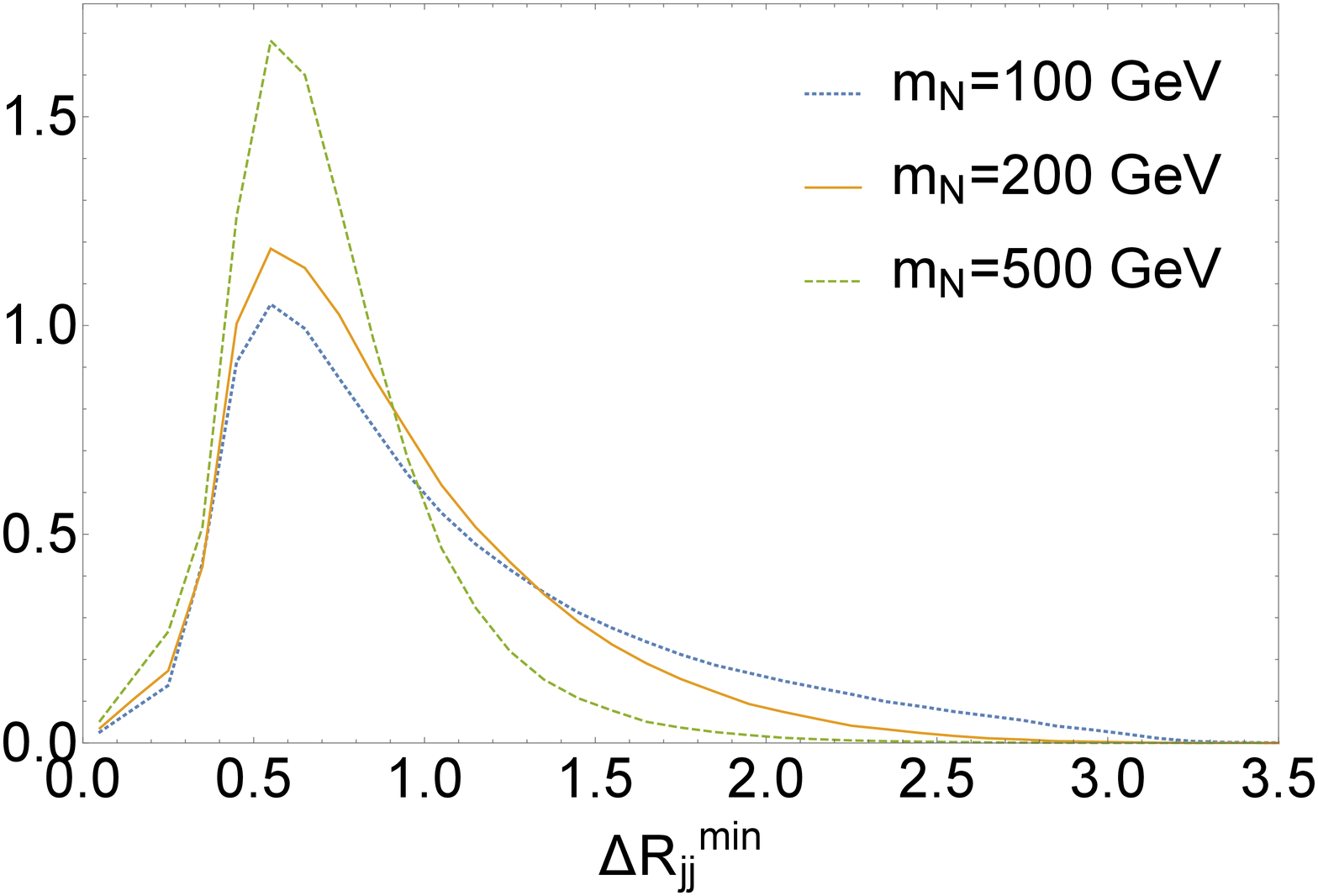}&
  \includegraphics[width=0.45\textwidth,clip]{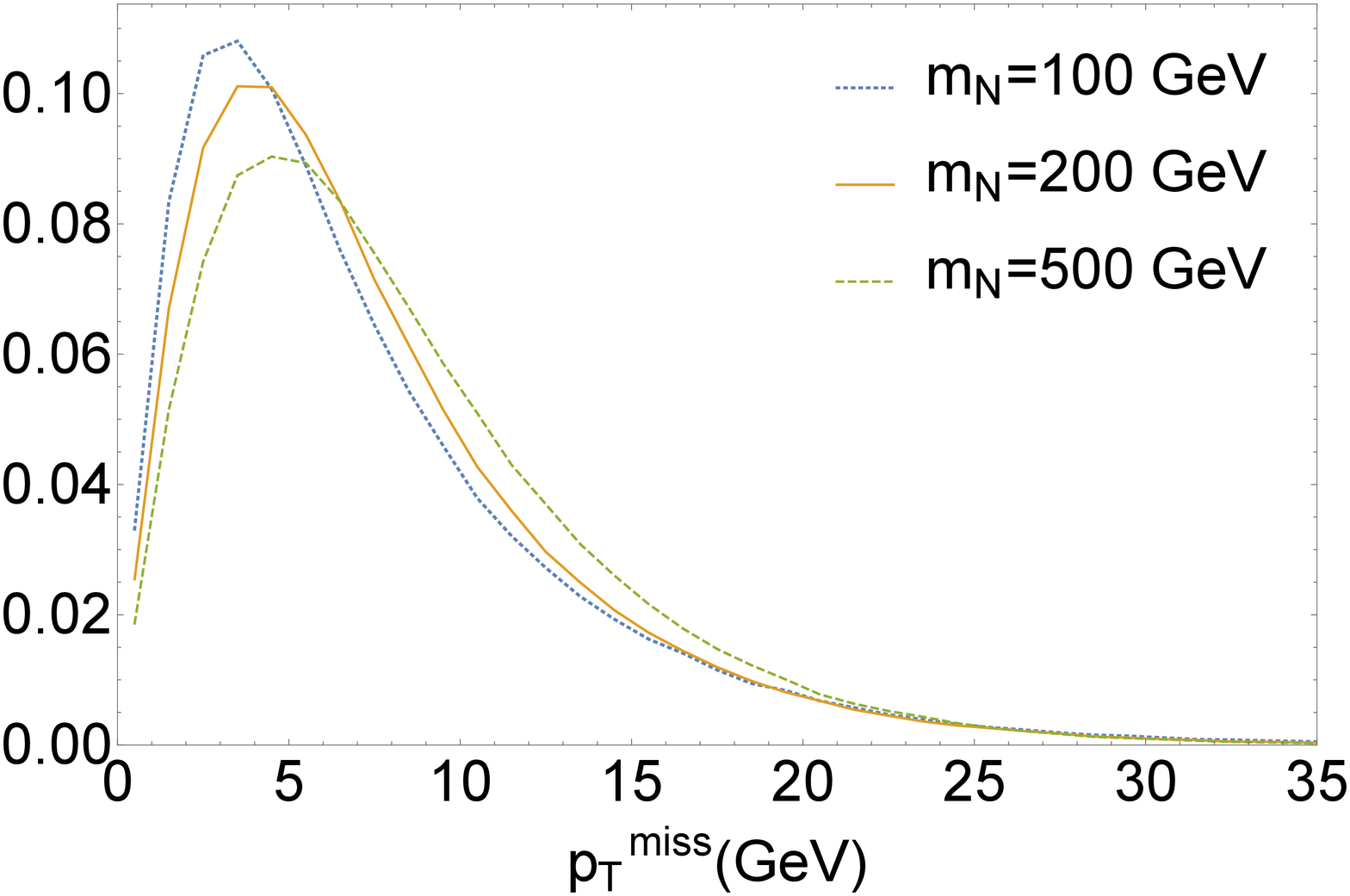}\\
  (c)&(d)\\
  \includegraphics[width=0.45\textwidth,clip]{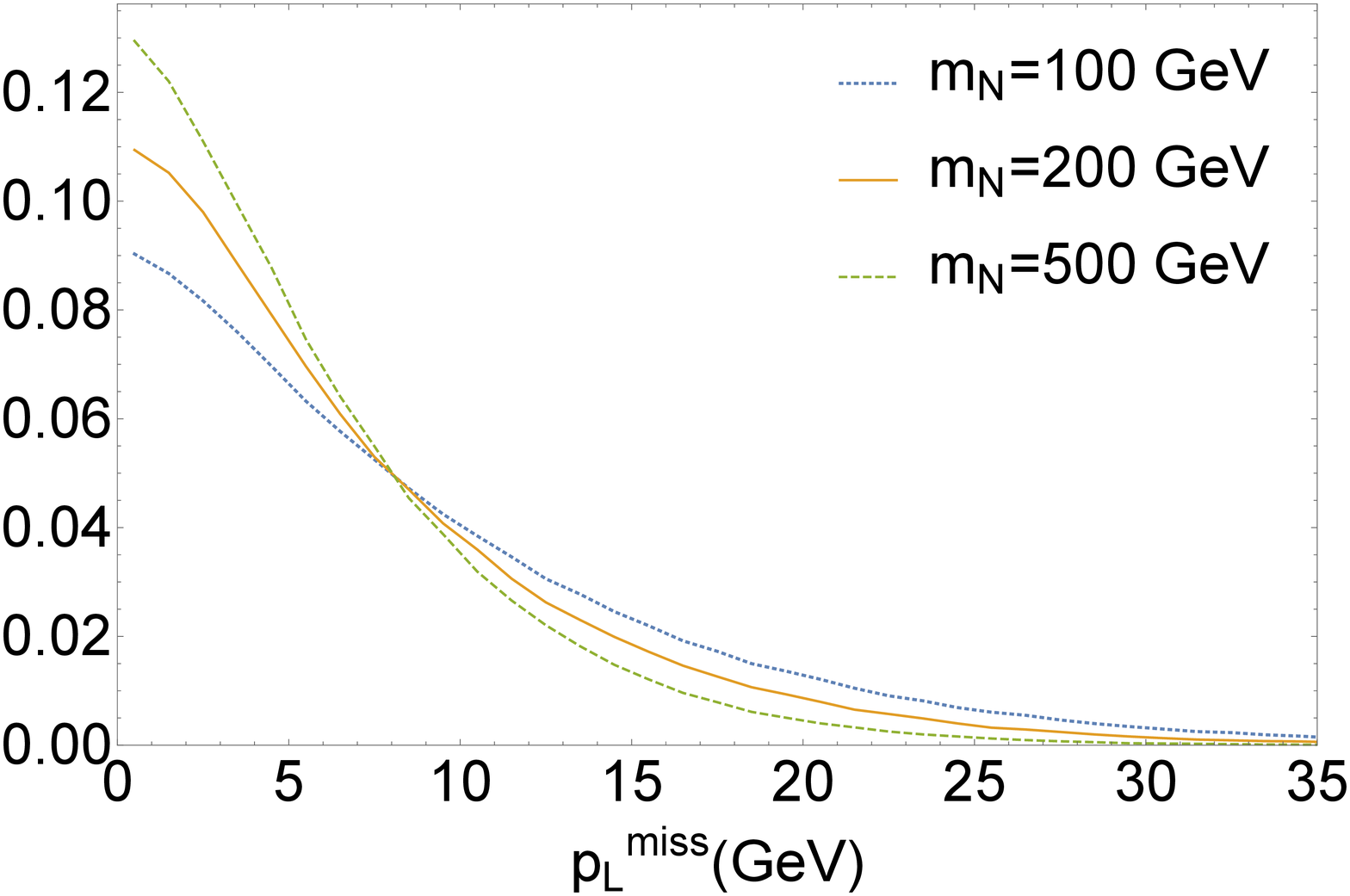}&
  \includegraphics[width=0.45\textwidth,clip]{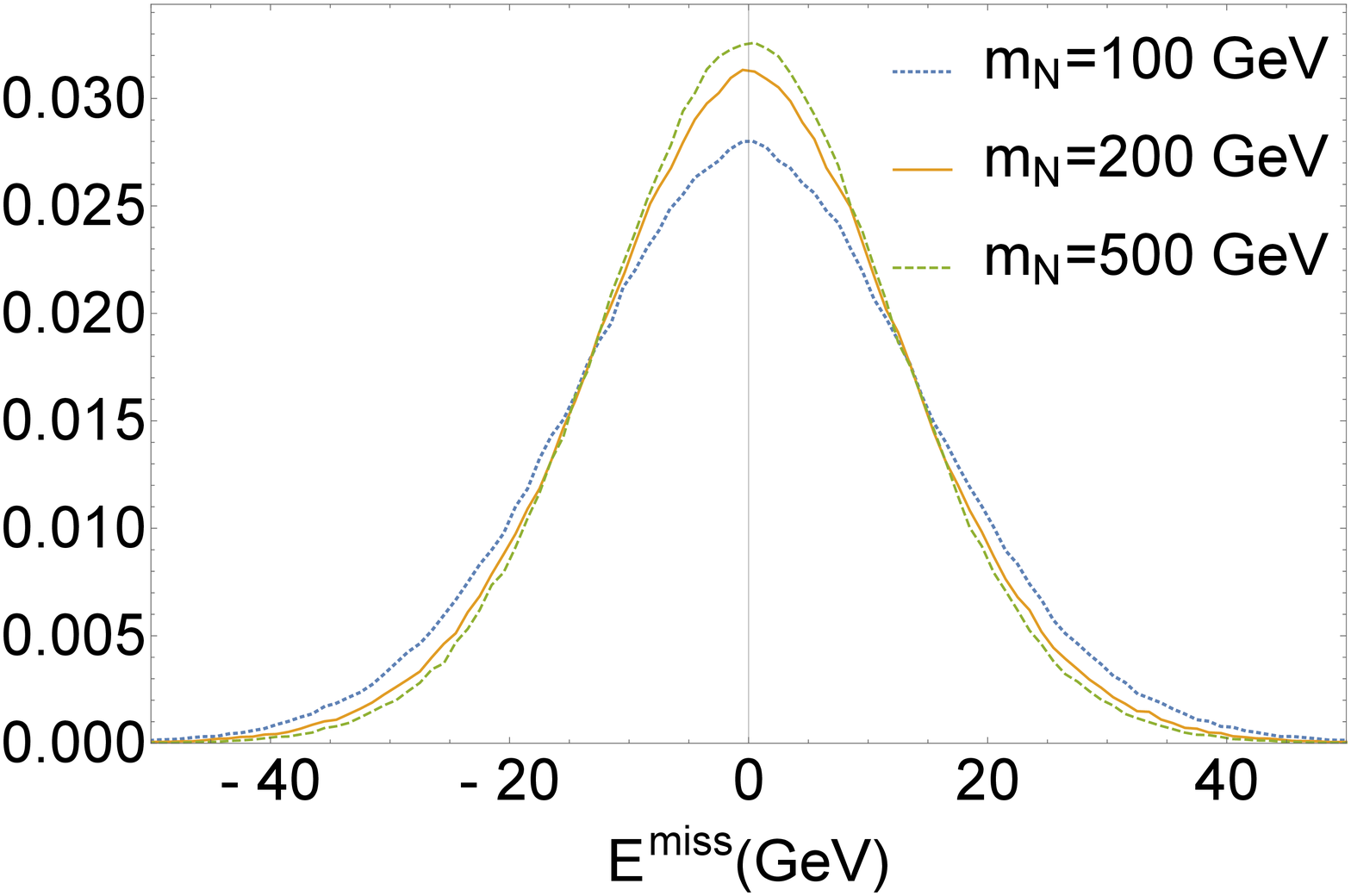}\\
  (e)&(f)
  \end{tabular}
  \caption{Normalized distributions $\sigma^{-1}\ud\sigma/\ud X$ for $m_N=100$, 200 and 500 GeV at the 1 TeV ILC for (a) the radial isolation of muon pairs $\Delta R_{\ell\ell}$; the minimal radial isolation (b) $\Delta R_{\ell j}^{\rm min}$ and (c) $\Delta R_{jj}^{\rm min}$; (d) the missing transverse momentum $p\!\!\!\slash_T$; (e) the missing longitudinal momentum $p\!\!\!\slash_L$ and (f) the missing total energy $E\!\!\!\!\slash$. The missing total energy is c.m. energy 1 TeV GeV minus the sum of energies of all visible particles in the final state.}
  \label{fig:ilcspc}
\end{figure}

At 1 TeV c.m. energy, a significant background comes from four on-shell $W$ bosons production, with two like-sign ones decaying leptonically and the other two decaying hadronically
\begin{equation}
e^- e^+\rightarrow W^+W^+W^-W^-\rightarrow \mu^\pm\mu^\pm\nu\nu+4j
\end{equation}
The same final state can be produced via the process
\begin{equation}
e^- e^+\rightarrow W^\pm W^\pm W^\mp jj\rightarrow \mu^\pm\mu^\pm\nu\nu+4j
\end{equation}
We also analysed other possibilities of four on-shell gauge bosons production, for example
\begin{equation}
e^- e^+\rightarrow ZZW^+W^-\rightarrow \mu^\pm\mu^\pm\mu^\mp\nu+4j
\end{equation}
Similarly, because of two missing leptons (neutrinos or charged muons), these channels can be effectively suppressed by appropriate cuts on missing energy. As an illustration, we plot the distributions of $p\!\!\!\slash_T$, $p\!\!\!\slash_L$ and $E\!\!\!\!\slash$ of the four $W$ bosons channel in Fig.~\ref{fig:mpilc4w}.
\begin{figure}[tb]
  \centering
  \includegraphics[width=0.5\textwidth,clip]{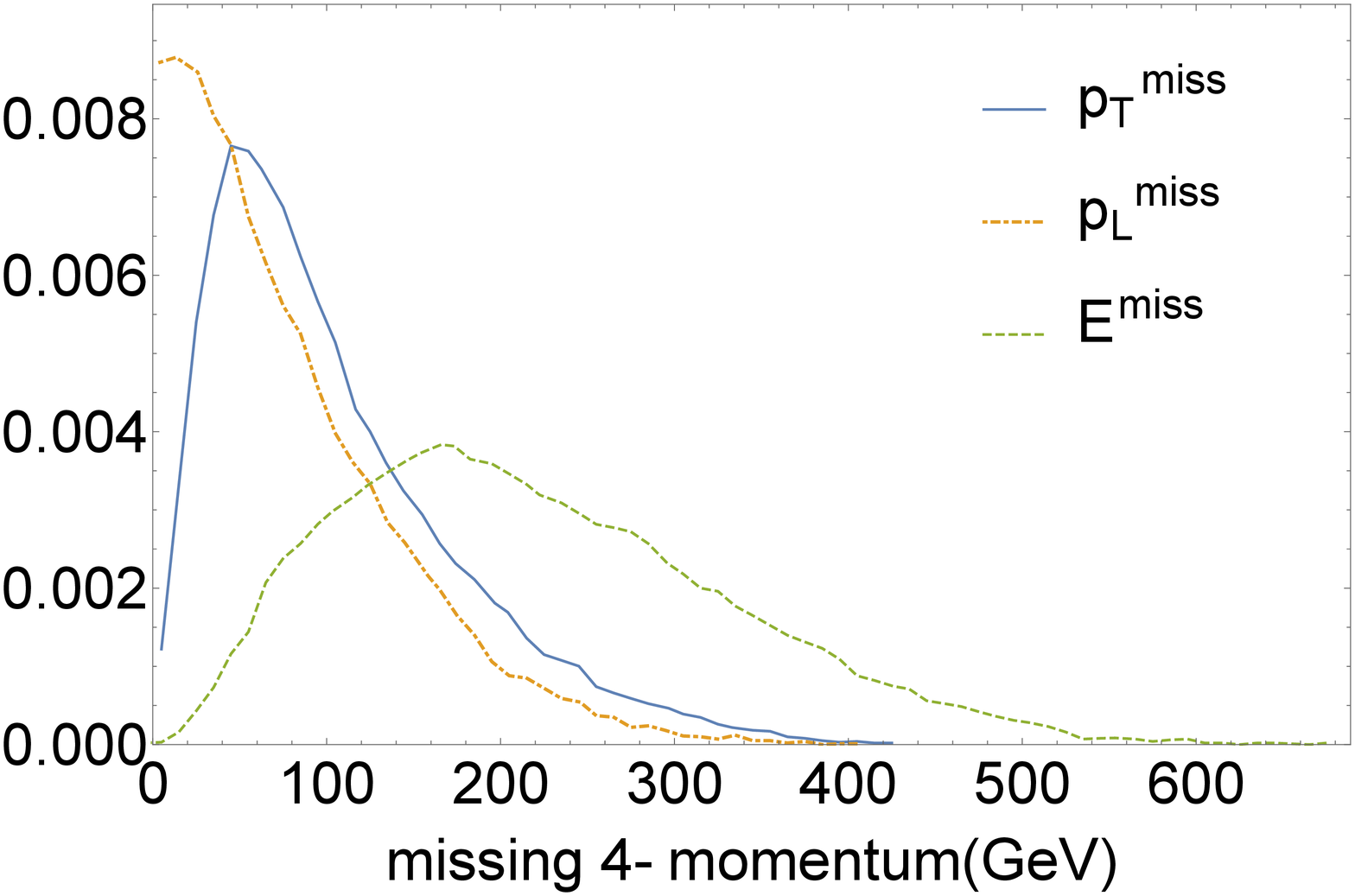}
  \caption{Normalized distributions $\sigma^{-1}\ud\sigma/\ud X$ for background channel $e^- e^+\rightarrow W^+W^+W^-W^-\rightarrow \mu^\pm\mu^\pm\nu\nu+4j$ at the ILC for the missing transverse momentum $p\!\!\!\slash_T$, missing longitudinal momentum $p\!\!\!\slash_L$ and missing total energy $E\!\!\!\!\slash$. The missing total energy is c.m. energy 1000 GeV minus the sum of energies of all visible particles in the final state.}
  \label{fig:mpilc4w}
\end{figure}

According to the above spectrum analysis, beyond the basic acceptance cuts, we impose the following additional cuts on radial distance and missing four-momentum to guarantee qualified events:
\begin{align}
 & \Delta R_{\ell\ell}>0.4,\quad \Delta R_{\ell j}^{\rm min}>0.4,\quad \Delta R_{jj}^{\rm min}>0.4,\label{cut7}\\
 & p\!\!\!\slash_T < 20~\gev,\quad E\!\!\!\!\slash < 40~\gev.\label{cut8}
\end{align}
After implementing all these cuts in Eqs.~(\ref{cut5})-(\ref{cut6}) and Eqs.~(\ref{cut7})-(\ref{cut8}), the SM background is essentially eliminated and no remaining events survive for the expected luminosity of 1000 $\ifb$ at the ILC. The signal cross sections for different heavy neutrino masses at the ILC are shown in Fig.~\ref{fig:cscut1000}, where we set $S_{\mu\mu}=10^{-4}$ as a benchmark value.
\begin{figure}[tb]
\begin{center}
  \includegraphics[width=0.5\textwidth,clip=true]{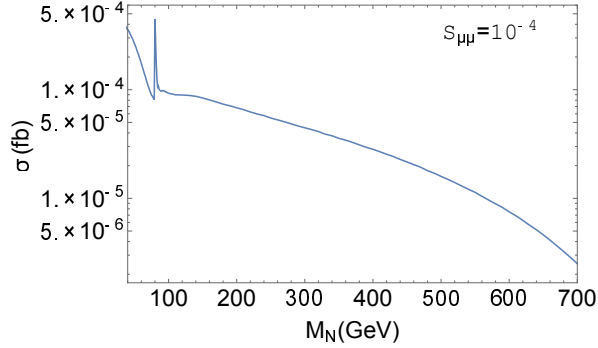}
  \caption{Total cross section for the process $e^+e^-\rightarrow N\mu^\pm W^\mp\rightarrow\mu^\pm\mu^\pm+4j$ after trigger cuts at the ILC, with $S_{\mu\mu}=10^{-4}$.}
  \label{fig:cscut1000}
\end{center}
\end{figure}
The reduction in rate is mainly due to basic acceptance cuts, and the cross section gradually drops with increasing $m_N$.

Again we adopt Poisson statistics to determine signal significance. Our final results about the sensitivity on the mixing parameter $S_{\mu\mu}$ are shown in Fig.~\ref{fig:excline1000}, compared to previous results from the LEP experiments \cite{Adriani:1992pq} and the LHC simulations in \cite{Atre:2009rg}.
\begin{figure}[!htbp]
  \centering
  \includegraphics[width=0.5\textwidth,clip]{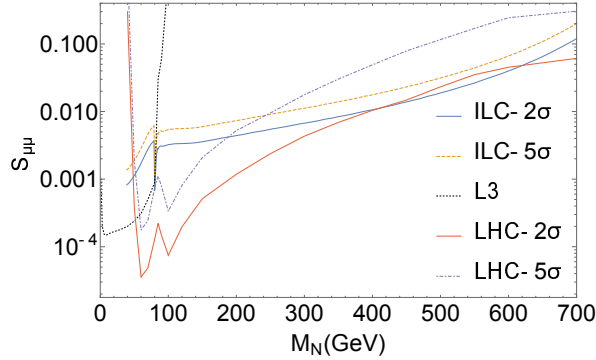}
  \caption{Sensitivity on $S_{\mu\mu}$ at the ILC. For comparison, the 95\% bound from L3 search and simulated LHC results in \cite{Atre:2009rg} are presented.}
  \label{fig:excline1000}
\end{figure}
We see that compared to previous results, the ILC can improve measurement sensitivity on mixing parameter $S_{\mu\mu}$ for more massive Majorana neutrinos. At a $5\sigma$ level, it improves sensitivity by about one order of magnitude for mass between 350 and 600 GeV. At a $2\sigma$ level, there is also a slight amount of improvement in the mass range of $400-600$ GeV. It is worth mentioning that since the $W\gamma$ fusion channel mentioned above already makes significant contribution in this mass range at the 14 TeV LHC, if this channel as well as the next-to-next-to-leading order (NNLO) QCD corrections are considered, the advantages of the ILC will become weaker and may even vanish \cite{Dev:2013wba,Alva:2014gxa}.

\section{Summary and Conclusions}

In this work, we  study a production channel of Majorana neutrinos at lepton colliders which violate lepton number by two units ($\Delta L=2$): $e^+e^-\rightarrow N\mu^\pm W^\mp\rightarrow\mu^\pm\mu^\pm+4j$. This process is of profound implication for understanding the nature of neutrino masses. Although the cross section of signal process is relatively small, lepton colliders still have rather ability to search for  signals of  Majorana neutrinos due to the clean environment and high integrated luminosity of the next generation lepton colliders. At the CEPC, the detection limit on mixing parameter $S_{\mu\mu}$ can be probed to about $3\times10^{-6}$ at a $2\sigma$ level and $7\times10^{-6}$ at a $5\sigma$ level in the neutrino mass range of $30-50$ GeV. While at the ILC which has a comparative advantage in larger mass range, it improves measurement sensitivity on $S_{\mu\mu}$ by about one order of magnitude in the mass range of $350-600$ GeV at a $5\sigma$ level. In summary, both for less and more massive Majorana neutrinos, future generation lepton colliders have the potential to make progress on searching for   Majorana neutrinos.

\bibliographystyle{JHEP}
\end{document}